\pdfoutput=1
\documentclass[preprint,5p]{elsarticle}
\usepackage[table,dvipsnames]{xcolor}
\usepackage{amsmath,amssymb}
\usepackage{graphicx}
\usepackage{hyperref}

\graphicspath{{./}{./}}

\makeatletter
\newcommand\footnoteref[1]{\protected@xdef\@thefnmark{\ref{#1}}\@footnotemark}
\makeatother

\newcommand{\vivo}{\textit{in-vivo }}

\newcommand{\C}{$\mathcal{C}$ }
\newcommand{\CTE}{$\mathcal{CTE}$ }
\newcommand{\y}{\overline{y}}
\newcommand{\yi}{\y(i)}

\begin{document}
\begin{frontmatter}

\title{
Predicting Chemical Hazard across Taxa through Machine Learning
}
\author[1,2]{Jimeng Wu\fnref{fn1}}
\ead{jimeng.wu@eawag.ch}
\author[3]{Simone D'Ambrosi\fnref{fn1}}
\author[4]{Lorenz Ammann}
\author[1]{Julita Stadnicka-Michalak}
\author[1,5]{Kristin Schirmer}
\ead{kristin.schirmer@eawag.ch}
\author[1]{Marco Baity-Jesi\corref{cor1}}
\ead{marco.baityjesi@eawag.ch}

\fntext[fn1]{Equal contribution.}
\cortext[cor1]{Corresponding author.}

\address[1]{Eawag, \"Uberlandstrasse 133, CH-8600 D\"ubendorf, Switzerland}
\address[2]{Department of Environmental Engineering, ETHZ, Zurich, Switzerland}
\address[3]{Department of Statistics, Sapienza University of Rome, Piazzale Aldo Moro, 5, 00185 Rome, RM, Italy}
\address[4]{Swiss Federal Institute for Forest, Snow, and Landscape Research WSL, Z\"urcherstrasse 111, CH-8903 Birmensdorf, Switzerland}
\address[5]{School of Architecture, Civil and Environmental Engineering, EPFL, Lausanne, Switzerland}

\date{\today}

\begin{abstract}
We applied machine learning methods to predict chemical hazards focusing on fish acute toxicity across taxa. We analyzed the relevance of taxonomy and experimental setup, showing that taking them into account can lead to considerable improvements in the classification performance. 
We quantified the gain obtained throught the introduction of taxonomic and experimental information, compared to classification based on chemical information alone. 
We used our approach with standard machine learning models (K-nearest neighbors, random forests and deep neural networks), as well as the recently proposed Read-Across Structure Activity Relationship (RASAR) models, which were very successful in predicting chemical hazards to mammals based on chemical similarity. We were able to obtain accuracies of over 93\% on datasets where, due to noise in the data, the maximum achievable accuracy was expected to be below 96\%. The best performances were obtained by random forests and RASAR models. 
We analyzed metrics to compare our results with animal test reproducibility, and despite most of our models 
 “outperform animal test reproducibility” as measured through recently proposed metrics, we showed that the comparison between machine learning performance and animal test reproducibility should be addressed with particular care.
While we focused on fish mortality, our approach, provided that the right data is available, is valid for any combination of chemicals, effects and taxa.
\end{abstract}
\end{frontmatter}

\section{Introduction}
One of the pillars of modern civilization is the ability to synthesize and/or use an enormous range of chemicals, which allow for new and improved products and serve as pharmaceuticals, pesticides, food additives and the like. However, their benefits need to be weighed against their risk. 
For that purpose, governments put risk assessment procedures in place decades ago, with the aim of evaluating the impact of chemicals on human and environmental health. In these assessment procedures, exposure, \textit{i.e.} the level of chemical to which organisms are subjected, and hazard, \textit{i.e.} the ability of chemicals to cause negative effects on organism/population health, are considered together as determinants of risk~\cite{setac:18}. Hazard assessment traditionally depends, to a large part, on animal tests using vertebrates, such as tests with mice, rats and fish, depending if assessments are for human or environmental health. These animal tests are ethically controversial. They are
also highly resource consuming (time, personnel, test material needed) and can, in fact, not comply with the demand of rapid testing of an ever increasing chemical universe in need of evaluation~\cite{rovida:09}. On these grounds, there are global efforts to refine hazard assessments using alternative or supplementary strategies~\cite{hamm:17,sullivan:21}. One of them is the employment of machine learning~\cite{vo:19}.

Machine learning (ML) is a term broadly applied to the use of computational algorithms to infer emerging patterns from data. 
In the field of chemical hazard assessment, one can mine chemical structural and physico-chemical information, along with effect data reported from animal toxicity tests. This comes often with the advantage that the domain of applicability is unrestricted, provided that appropriate training data is available. 

One of the first applications of ML to hazard assessment was the prediction of acute oral toxicity in rats (LD50 values), using local lazy learning, achieving correlation coefficients of $R^2=0.72$ on a test set of $2896$ compounds~\cite{lu:14}. 
Further ML approaches to predict rat acute oral toxicity by other investigators~\cite{wu:18} improved the $R^2$ to 0.86, and using discrete outputs to measure accuracies, values reached close to 95\%~\cite{xu:17}.
However, with their focus on single species data, these studies do not cover the dimensionality of taxa and species sensitivity differences to be considered if one wants to know chemical impacts on another species, such as humans, or on the many species constituting an ecosystem~\cite{furuhama:15,basant:16,ai:19}, the latter of which is the focus of ecotoxicology. 
In this case, the use of ML is even more crucial, since in practice it is unthinkable to test all chemicals on all the species present in an ecosystem.
% A limitation that is circum
Acute toxicity predictions have also been the focus of few applications of ML in \emph{eco}toxicology, though a critical lack of literature on machine learning in this area has been denounced~\cite{miller:18}. 
Li \textit{et al.} aimed to predict toxicity categories for acute toxicity to fish of pesticides by making use of nine chemical descriptors~\cite{li:17b}. Their dataset was rather small ($\sim1000$ entries), and dominated by two species, rainbow trout (RT) and bluegill sunfish (BS). They trained models on the RT and BS data separately, and also trained models by mixing all the available taxa together. In binary classification, %\textit{i.e.}, 
their balanced accuracy (average between sensitivity and specificity, see Sec.~\ref{sec:metrics}) was lowest when excluding RT and BS, it was about 0.815 when restricting to a single taxon, and 0.825 when treating the whole dataset. However, it is difficult to state whether these differences are statistically significant, and, if so, {whether it should be attributed to the smaller size of the dataset when training without the RT and BS.}
Tuulaikhuu \textit{et al.} studied the EcoTox database, which contained enough data to allow for using taxonomy as a training feature~\cite{tuulaikhuu:17}. They trained random forests to predict acute fish toxicity data (LC50), obtaining an $R^2=0.85$. They identified fish species and the octanol-water partition coefficient (logP or logK$_\mathrm{ow}$) as the most important factors to explain species sensitivities differences. Despite the lack of a hold-out test set (in addition to train and validation), this result seems to suggest that in ecotoxicology we should be training machine learning models that infer how toxicity varies across taxa.

Indeed, when measuring the performances of different approaches, one of the fundamental questions is whether the measured differences are statistically significant. %This issue was taken upon by Hou \textit{et al.}~\cite{hou:20}, by taking 10 independent hold-out validation sets, and taking average among them. 
{This issue was tackled by Hou \textit{et al.}~\cite{hou:20}, by taking an average of 10 independent hold-out validation sets. }This procedure allowed an estimation of statistical fluctuations, but it also imposed a burden on the training data availability, since these hold-out test sets cannot be used for training. 

Another fundamental question is how good should we expect machine learning models to be, since \textit{in-vivo} toxicity data is often noisy
(\textit{i.e.} repetitions some of the same experiment can provide very different results), and testing on noisy data cannot return a perfect performance.
This issue was inspected by Luechtefeld \emph{et al.}~\cite{luechtefeld:18}, who, focusing on specific OECD\footnote{Organisation for Economic Co-operation and Development.} chemical testing guidelines %OECD test guidelines 
(all concerned with rabbits, guinea pigs, mice, rats or Chinese hamster), adopted a metric to compare the performance of their machine learning model with the consistency of the experiments themselves. Through this metric, they concluded that their model, called Read-Across Structure Activity Relationship (RASAR), is "outperforming animal test reproducibility".
RASAR models were developed for binary classification, and adopt a succession of unsupervised and supervised learning, which allows the exploitation of information on endpoints that are different from the target one. 
{For example, one could predict the LC50 of a chemical by integrating a Mortality with a Behavior dataset.} Given the great success of RASAR models, a natural question is whether they perform equally good in the datasets used in ecotoxicology, and how they compare to standard ML models.

Here, we perform an extensive analysis of ML models applied to an ecotoxicological dataset focused on fish acute toxicity data, with an eye on the extra information that can be obtained from the knowledge about the toxicity of the same compound on more than one species. 
We compare several standard machine learning models with the two variants of the RASAR model developed by
Luechtefeld \textit{et al.}~\cite{luechtefeld:18}. 
We extend RASAR models to an arbitrary number of classes, which allows us to study both binary (more toxic versus less toxic) and 5-class (non-toxic, slightly toxic, moderately toxic, highly toxic, very highly toxic) classification. 
We carefully analyze the impact of introducing taxonomic and experimental information as extra input features in addition to the chemical, since this introduction allows for models with a wide taxonomic range of applicability, which can learn the patterns relating to toxicity and taxonomic/experimental variability.
Finally, we compare our results with two indicators of the reproducibility of the experiments. These suggest that, given the level of noise in \textit{in-vivo} data (\textit{e.g.} the same experiment repeated twice may have very different outcomes), we are close to the maximum achievable performances. However, this should not be taken as an indication that these models are outperforming animal test reproducibility.

\section{Materials and Methods}

\subsection{Problem Setup}\label{sec:notation}
We are interested in predicting whether an experiment will result in a certain chemical being more or less acutely toxic on a given species. The outcome of the experiment is the LC50, which is the concentration of a chemical required to kill half a population, in our case 50\% of fish in a 24h, 48h, 72h or 96h acute exposure scenario. We treat the problem as binary classification according to the European regulation (EC) No 1272/2008 on the classification, labelling and packaging of substances and mixtures (CLP Regulation) for acute toxicity data~\cite{reach-binary}. To do so, we define the binary target label $y_2$:
\begin{equation}\label{eq:binary}
    y_2 = \begin{cases}
    1 ~~~\text{if~~LC50}  \leq1\text{mg}/\ell \,,\\
    0 ~~~\text{if~~LC50}  > 1\text{mg}/\ell\,.
    \end{cases}
\end{equation}

Our main goal is to deduce $y_2$ from the data, which we denote as $\vec x$.
 In \ref{app:multiclass} we show analogous results for five-class labels.
 
\subsection{Data}\label{sec:data}
The data, $\vec x$, contains information on chemicals, $\vec x_\mathrm{ch}$, on the taxonomy of the tested organism, $\vec x_\mathrm{tax}$, and on the experimental conditions, $\vec x_\mathrm{ex}$:
\begin{equation}
    \vec x = (\vec x_\mathrm{ch}, \vec x_\mathrm{tax}, \vec x_\mathrm{ex})\,.
\end{equation}
The vectors $\vec x_\mathrm{ch}, \vec x_\mathrm{tax}$ and $\vec x_\mathrm{ex}$ have respectively 893, 5 and 6 components.
To build $x$ and the related set of labels [Eq.~\eqref{eq:binary}], we take \vivo experiments on fishes from the Ecotox database (\href{https://cfpub.epa.gov/ecotox/}{https://cfpub.epa.gov/ecotox/}), provided by the Environmental Protection Agency of the USA. This database contains $\vec x_\mathrm{tax}$ and $\vec x_\mathrm{ex}$, but it only contains the CAS identifier for the used chemical.
In \ref{app:data} we describe in detail the sources of the data, and which features are contained in $\vec x_\mathrm{ch}$, $\vec x_\mathrm{tax}$, and $\vec x_\mathrm{ex}$.\\
The used information on the chemical has two different natures. We use some standard properties of each chemical (\textit{e.g.} logP, water solubility, etc.); and the Pubchem2D vectors, that identify chemicals through a one-hot encoded 881-dimensional vector (see \ref{app:data}). In other words, we can write $\vec x_\mathrm{ch}$ as the union of the properties $\vec x_\mathrm{pr}$ and the Pubchem2D data $\vec x_\mathrm{pubc}$,
\begin{equation}
    \vec x_\mathrm{ch} = (\vec x_\mathrm{pr}, \vec x_\mathrm{pubc})\,.
\end{equation}

We study \vivo mortality experiments performed on fishes, and take into account all the chemicals (mainly organic compounds) and species available in the dataset. As for the taxonomic ranks, we use class, order, family, genus and species as features. 
When the same experiment is repeated more than once, we merge the multiple entries into a single one, and take the median LC50 as an outcome.
The data cleaning is described %as well 
in \ref{app:data}.

\begin{figure}[tb]
    \centering
    \begin{tabular}{@{}cc@{}}
    \includegraphics[width=0.45\linewidth]{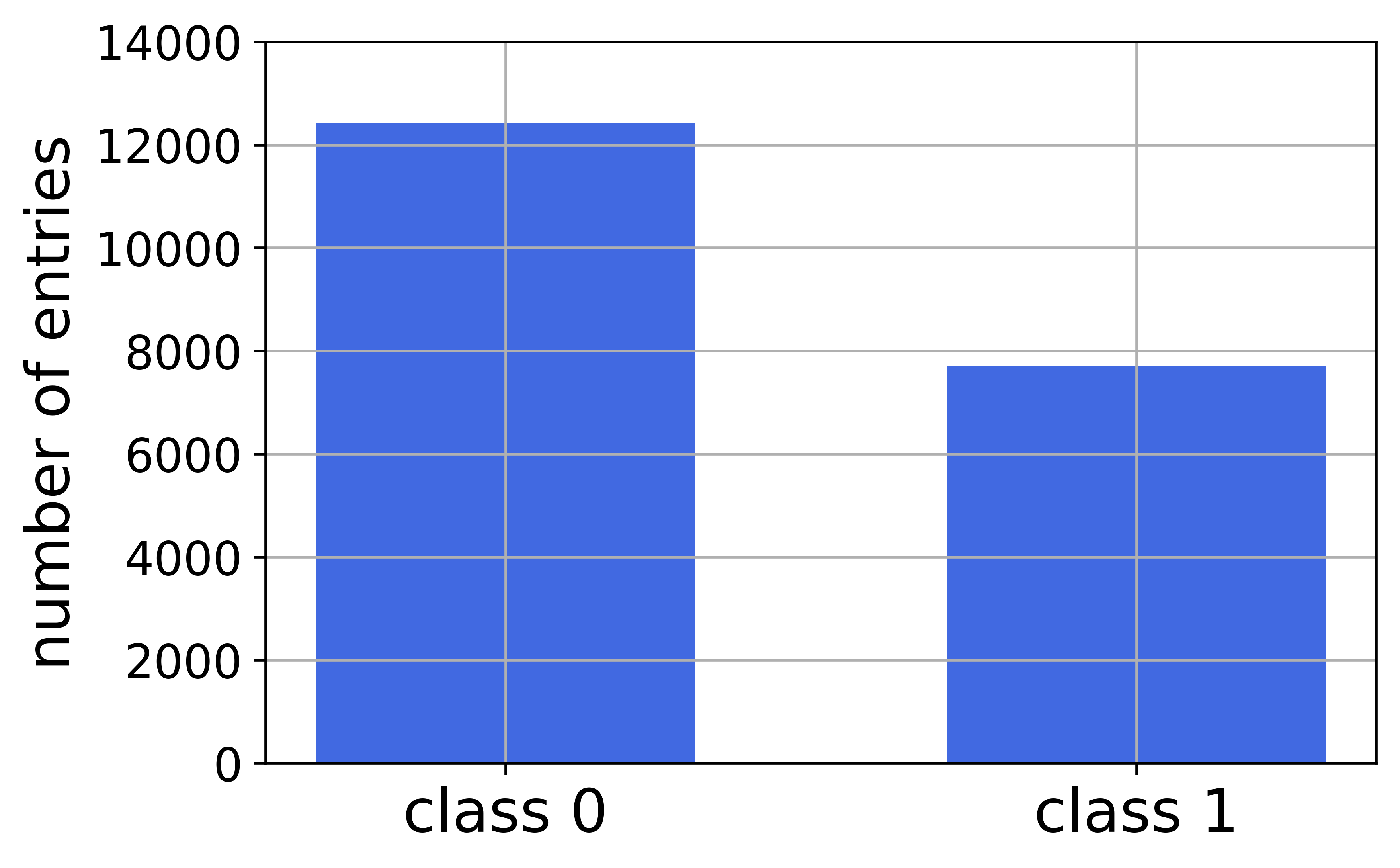}
    \includegraphics[width=0.45\linewidth]{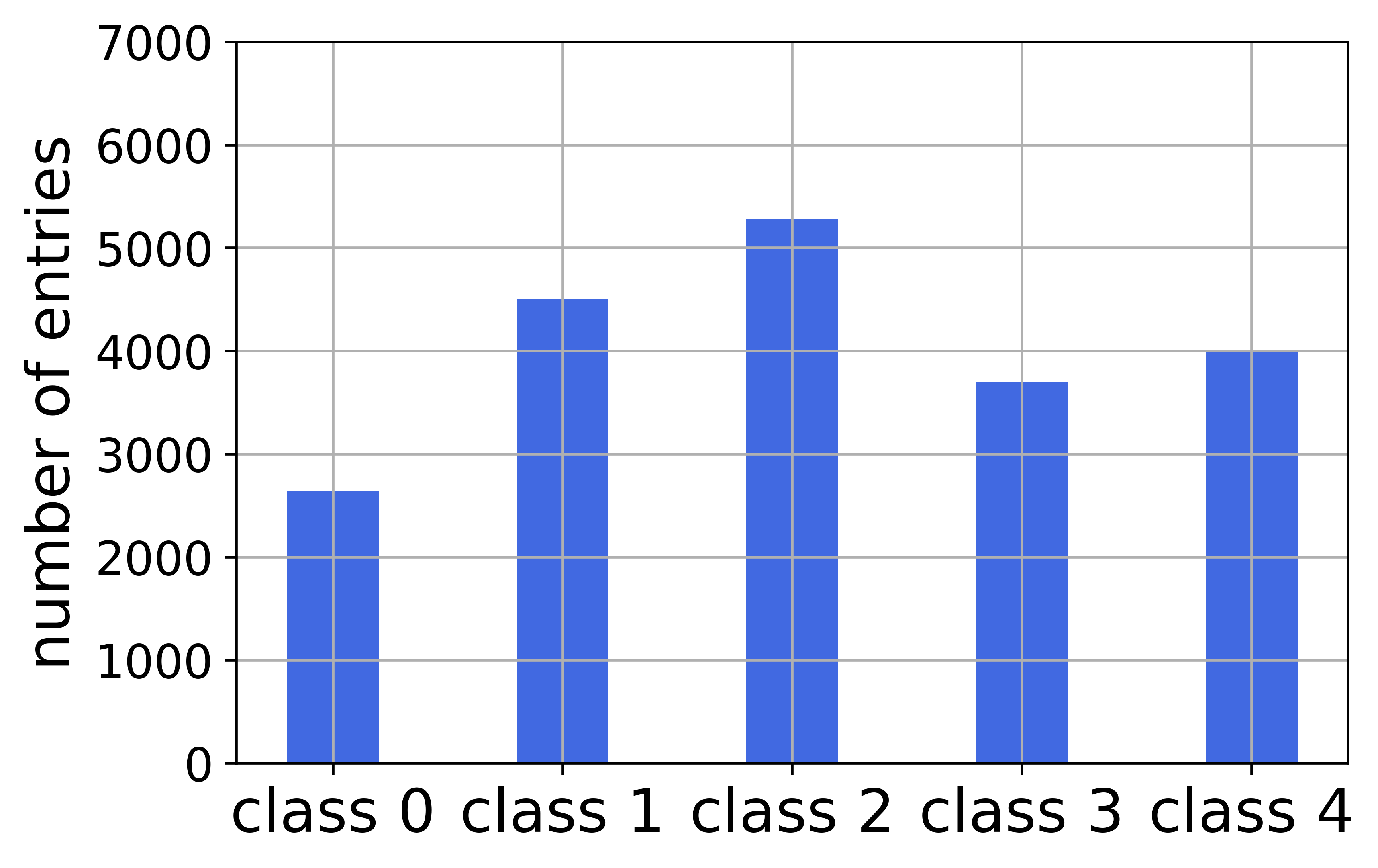}
    \end{tabular}
    \caption[class proportions]{Number of experiments per label after binary and 5-class labeling of the concentrations. The thresholds that define the classes are summarized in the appendix (Tab.~\ref{tab:classes}).}
    \label{fig:class}
\end{figure}

At the end of the process, we have $n_\mathrm{chem}=2199$ chemicals, tested for mortality on $n_\mathrm{spec}=345$ different species of fish. In total, our dataset counts 20128 entries.
In Fig.~\ref{fig:class}, we show how the data is distributed into each class, in the case that we split the outputs into 2 (left) or 5 classes (right).

In addition to mortality, we also use data related to different effects, which is necessary as an input for the Data-Fusion RASAR model (\ref{app:models}).
This adds 3713 further entries to our dataset, spanning over 543 chemicals. Of those, 117 do not appear in the dataset related to mortality. See \ref{app:data} for details.

% The ecotoxicological nature of this work stands out from the large number of different fish species in the dataset, and from the fact that we aim at gaining information from all those species, and using it to better predict the toxicity on other species. 

%This is different from what we call a \emph{toxicological approach}, where machine learning models dismiss the information on the species, either by only focusing on a single species, or assuming that a chemical acts similarly on different species. 

\subsection{Models}\label{sec:models}
We train Logistic Regressions (LR), Random Forests (RF), K-Nearest-Neighbors (KNN), Multi-layer Perceptrons (MLP, this is a standard deep neural network model)~\cite{alpaydin:20,goodfellow:16}, Simple RASAR (S-RASAR) and Data Fusion RASAR (DF-RASAR)~\cite{luechtefeld:18}. Details on the models are provided in \ref{app:models}.

\subsubsection{Euclidean Distances}
The KNN and RASAR algorithms require measuring a distance between feature vectors. Initially, we will use Euclidean distances. In other words, the distance between two feature vectors $\vec u$ and $\vec v$ is
\begin{equation}
    \label{eq:naif-dist}
    d(\vec u,\vec v) = \sqrt{(\vec u - \vec v)^2}\,,
\end{equation}
regardless of whether these vectors have ordinal or one-hot components.

\subsubsection{Per-category Distances}\label{sec:def-alphas}
The fact that our data is clearly separated into the three sources of information, $x_\mathrm{ch},x_\mathrm{ex}$ and $x_\mathrm{tax}$, allows us to tailor our models to achieve better performance. One way to do this, is to define the distances in the KNN and RASAR models as a linear combination of the distances in different subspaces of the data.

We decide to treat separately the pubchem vectors (binary features), the chemical fingerprints (ordinal features), and the union of experimental and taxonomic details (these are categorical features, so we denote them with $\vec x_\mathrm{cat}=(\vec x_\mathrm{tax},\vec x_\mathrm{ex})$).

We measure the distance between pubchem vectors, through the Hamming distance function, $d_\mathrm{Ham}$,
% \footnote{We also tried to use the Tanimoto distance~\cite{cha:07}, but the Hamming distance turned out to perform better (see Appendix).}
which measures the number of equal entries in the two vectors. The distance between chemical fingerprints is a Euclidean distance, $d_\mathrm{Euc}$, and for the categorical features we also use the Hamming distance.
In order to have distances normalized to one (so that each set of features has the same weight), we divide the Hamming distances between two vectors by their number of components, and the Euclidean distances by their maximum over the training set. We call $\tilde{ d}_\mathrm{Ham}$ and $\tilde{d}_\mathrm{Euc}$ the resulting normalized distances.

The distance between two experiments $\vec u$ and $\vec v$ can then be written as
\begin{align}
\nonumber
    d(\vec u,\vec v) &= 
    \alpha_\mathrm{pr} \tilde{d}_\mathrm{Euc}(\vec u_\mathrm{pr},\vec v_\mathrm{pr}) \,+\\
\nonumber
    &+\alpha_\mathrm{pubc} \tilde{d}_\mathrm{Ham}(\vec u_\mathrm{pubc},\vec v_\mathrm{pubc}) +\,\\
    &+ \alpha_\mathrm{cat} \tilde{d}_\mathrm{Ham}(\vec u_\mathrm{cat},\vec v_\mathrm{cat})\,,
\label{eq:dist}
\end{align}
where $\alpha_\mathrm{pr},\alpha_\mathrm{pubc}$ and $\alpha_\mathrm{cat}$ are constants whose optimal value is found through cross validation (CV).

Since we are interested in the relative (and not absolute) values of these constants, we can safely set $\alpha_\mathrm{pr}=1$.
Then, to find the optimal values of $\alpha_\mathrm{pubc}$ and $\alpha_\mathrm{cat}$, we perform a search on a grid in the plane ($\alpha_\mathrm{pubc},\alpha_\mathrm{cat}$), tuning one dimension at a time, and choosing the alpha with the best accuracy. When two or more values of $\alpha$ give the same validation accuracy (with a tolerance of 0.001), we choose the smallest value. For both $\alpha_\mathrm{pubc}$ and $\alpha_\mathrm{cat}$, we first found the optimum a coarse logarithmic grid, and then fine-tuned it with a finer one.
Since we have several models that employ these distances, and because this hyperparameter tuning is computationally expensive, we only find the optimal $\alpha$s for the K-NN algorithm, and use those values for the S and DF RASARs.

The advantage of using the distances in Eq.~\eqref{eq:dist} is twofold:
\begin{enumerate}   
    \item It allows to assign a different weight to different subsets of features. This is useful because we do not expect all the groups of features to be equally important.
    Splitting the features into smaller groups can further improve the performances, but this would be at the expense of a longer hyperparameter tuning.
\item Since the distances $d_\mathrm{Euc},d_\mathrm{pubc}$ and $d_\mathrm{cat}$ are normalized, the constants $\alpha$ automatically tell us the relative importance of each source of data. In particular, the value of $\alpha_\mathrm{cat}$ gives us a direct indication of the role played by taxonomy and experiment details.
\end{enumerate}

\subsection{Training and Assessing Performance}
\label{sec:metrics}
We divide our dataset in a training and a hold-out test set with an 80:20 ratio. 
Unless stated otherwise, the training set is further split into train and validation sets, through a 5-fold cross-validation procedure (the splitting is again 80:20)~\cite{arlot:10}. After we found the best model hyperparameters through an exhaustive grid search during CV, we retrain the model on the whole train+validation set. The final performances are calculated on the test set.
Except when explicitly stated, we do not use methods to mitigate class imbalance.

We measure the performances in terms of accuracy ($a$), recall ($r$, often called sensitivity), specificity ($s$), precision ($p$) and F1 score (F$_1$). {The definition of these indicators is provided in \ref{app:metrics}.}

\subsection{Error estimates}\label{sec:errors}
When possible, in order to get an intuition on whether two results are significantly different, we compute error bars on the performances as the standard error between the cross-validation folds. 
These error bars represent the performance fluctuations of models trained only on the validation folds. Since the validation folds are smaller than the test set, the fluctuations on the test set should be smaller. In addition, while the performances on the validation folds are computed on models trained on the training folds, the performance on the test set comes from a model trained on training+validation data, so we can expect the test performance to fluctuate less or equally.
While not rigorous, this procedure allows for an easier interpretation of the results.
The errors bars are the number given in parenthesis and are applied to the last significant digit (\textit{e.g.} 0.03(1) stands for 0.03$\pm$0.01).

\subsection{Reproducibility of the experiments}
\label{sec:repeated-exp}
Even though from the perspective of a regulator each experiment has one true outcome (in the binary case, more vs less toxic), if an experiment is performed more than once it sometimes happens that different repetitions of the experiment result in different values of the label $y_2$ (or $y_5$, if we are in the multiclass setting). We can interpret this variability as some source of fluctuations -- which derive from elements which are not completely under control -- around a \textit{ground truth} value.\footnote{An alternative would be to use a Bayesian approach, which describes the fluctuations as intrinsic, and would aim for example at reconstructing the full probability distribution of the LC50.}

Since these fluctuations are present in our whole data set, also the test set will have some degree of disagreement with the ground truth. As a consequence, testing a model that perfectly predicts the ground truth will result in an accuracy $\tilde{a}$ lower than $100\%$. This means that the maximum accuracy that we should expect for our models is not 100\%, but $\tilde{a}$, so the performances that we find should be measured in units of $\tilde{a}$. However, we do not have access to $\tilde{a}$, so the best we can to is to try to estimate it by analyzing the mutual (dis)agreement of experiments that were performed more than once.
We use two metrics: (a) an estimated upper bound to $\tilde{a}$, and (b) a conditional estimator proposed in Ref.~\cite{luechtefeld:18}.

\subsubsection{Upper Bound}
\paragraph{Intuitive definition}
We estimate an upper bound, $A_\mathrm{up}$, to the maximum achievable accuracy $\tilde{a}$. When all the repetitions of an experiment agree, we can assume that they correspond to the ground truth. When they do not coincide, the accuracy is maximized by assuming that the value of $y_2$ (or $y_5$) corresponding to the ground truth is the one occurring most often. We then define $A_\mathrm{up}$ as the accuracy obtained by assuming that repetitions of an experiment measure the ground truth more often than not. 
If we restrict the measurement of $A_\mathrm{up}$ to the experiments with a majority of positive outcomes, we obtain an upper bound to the recall, $R_\mathrm{up}$, whereas if we restrict it to the negative outcomes, we obtain an upper bound to the specificity, $S_\mathrm{up}$. The balanced accuracy is the average between $R_\mathrm{up}$ and $S_\mathrm{up}$. In \ref{app:formal-definitions} we provide the formal definitions, for the binary and multiclass cases.

\paragraph{An example}
To provide some intuition, let us focus on a single experiment ($n_\mathrm{rep}=1$), with binary labels, and assume that it is performed five times, with results $y_2 \equiv ({y_2}_1, {y_2}_2, {y_2}_3, {y_2}_4, {y_2}_5) = (0, 1, 0, 1, 1)$. Since the value $y_2=1$ appears most often, we assume that it corresponds to the ground truth, and obtain $A_\mathrm{up}=0.6$ (similarly, we could obtain a lower bound $A_\mathrm{down}=0.4$).
Notice that, by construction, $A_\mathrm{up}$ cannot be smaller than 0.5.

\subsubsection{Conditional Estimator}
\paragraph{Definition}
The conditional estimators of recall and specificity used in Ref.~\cite{luechtefeld:18} aim at calculating the actual value of those quantities, instead of an upper bound.
Their procedure postulates that one of the repetitions of the experiment is the true result, and validates the other repetitions assuming it as truth. For a formal definition we refer the reader to Ref.~\cite{luechtefeld:18}.

\paragraph{An example}
We can use the previously defined example of an experiment repeated five times, with results $y_2 = (0, 1, 0, 1, 1)$. We take the first outcome, ${y_2}_1=0$, as true, and see that the accuracy with the remaining experiments is $a_1=0.25$. We then iterate through all the outcomes, and obtain $\overline{a}=\frac15\sum_i^5{a_i}=0.4$.
For this specific example, this number is equal to the lower bound $A_\mathrm{down}$ obtained in the previous section, so it seems like a pessimistic estimate of the consistency of repeated experiments.
To estimate the recall $r$ (or specificity $s$), we restrict to the examples with a positive (or negative) outcome. We therefore have an estimate $r=0.5$ and $s=0.25$.

\paragraph{Experiment reproducibility metrics in our dataset}
In Tab.~\ref{tab:exp-acc} we show the value of the upper bound and conditional estimator measured from our data.

\paragraph{Caveat}
Note that these estimators $\tilde{a}$ are measured through different repetitions of the same experiment, whereas our models are trained and tested on the medians of these repetitions, which arguably make the labeling more stable. Thus, it is possible that $\tilde{a}$ is higher than what we would expect from our estimators.

\begin{table}[t]
\centering
\resizebox{\columnwidth}{!}{
\begin{tabular}{cc|cccc}
Metric      & Classification & Recall & Specificity & Bal. Acc. & Acc   \\ \hline \hline
Upper Bound & Binary         & 0.950  & 0.963       & 0.956     & 0.958 \\
Conditional & Binary         & 0.797  & 0.871       & 0.834     & 0.841 \\
Upper Bound & Multi          & 0.833  & -           & -         & 0.878     \\
Conditional & Multi          & 0.623  & -           & -         & 0.612    
\end{tabular}
}
\caption{Recall, specificity and balanced accuracy (average of recall and specificity) of repeated experiments, using two kinds of estimators defined in Section~\ref{sec:repeated-exp}.}
\label{tab:exp-acc}
\end{table}

\section{Results}\label{sec:res}

\subsection{Dealing with many species}
Compared with training and testing models on a single species, dealing with a large number of species requires a different perspective. On one side, [i] only a small number of species-chemical couples are present in {this kind of dataset.}
In other words, an ecotoxicological dataset is sparse. If our dataset was dense, it would have $n_\mathrm{spec}n_\mathrm{chem}=831222$ lines, instead of only 20128.
On the other side, [ii] the different chemicals have different degrees of toxicity depending on which species is exposed to them.

\begin{figure}
    \centering
    \includegraphics[width = 0.5\textwidth]{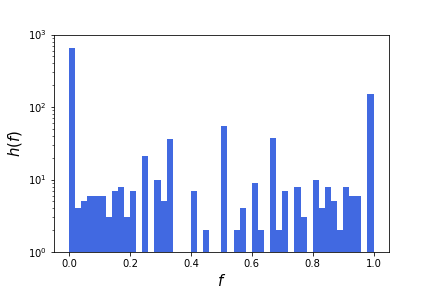}
    \caption{Histogram $h(f)$. For each chemical, the quantity $f$ is the fraction of species for which the given chemical is labeled as more toxic ($y_2=1$, after taking the median on the experiment repetitions as described in Sec.~\ref{sec:data}). If $f=0,1$, the chemical has the same binary outcome on all species, while if $f\neq0,1$, the chemical has a different impact according to the species.}
    \label{fig:hf}
\end{figure}

We can test statement [ii] from our dataset. For each chemical that is tested on more than one species, we measure the fraction $f$ of species for which it results more toxic (\textit{i.e.} it has a label $y_2=1$). If a chemical has the same effect on all taxa, then $f=0,1$.
We plot the histogram of $f$, $h(f)$, in Fig.~\ref{fig:hf}, where it can be seen that a sizable number of chemicals have $0<f<1$, meaning that these chemicals act differently on different species. The two peaks at 0 and 1 are expected, since we still expect that a number of chemicals have a similar effect across all the orders (especially because our analysis is restricted to a narrow taxonomic range, \textit{i.e.} fishes). Furthermore (i) many of the entries at $f=0,1$ come from chemicals which are tested on only a couple of species, and (ii) we did not filter $f$ according to the taxonomic distance between species. %If we restricted to only chemicals tested on a larger number of species, or repeated on sufficiently distant species, the peaks at $f=0,1$ would likely be lower.

The quantity $f$ fixes the chemical, and checks how different the outcomes are on different species. We can try instead to fix the taxon, in order to see whether the outcomes on a specific taxon are very different from the others.
To do this, for every taxon $t$, we take all the chemicals that were tested on $t$. Then, for each of these chemicals, we calculate the {mean value of} $f$, as defined in the previous paragraph. Finally, we calculate what we call the \textit{taxonomic disagreement} parameter,
\begin{equation}
    g=1-2\,|f-\frac12|,
\end{equation}
to indicate how different the outcomes on a taxon are from other taxa.
If, for taxon $t$, $g=0$, then all the chemicals that were tested on $t$ had the same outcome on all the other taxa on which they were tested. If $g>0$, it means that part of the chemicals that were tested on $t$ had a different outcome when tested on different values of the taxonomic rank.
In other words, if it was useless to use taxa in our dataset, we would always have $g=0$. The larger $g$ (the maximum value is 1), the more crucial it is to use the taxonomic information.

\begin{figure}
    \centering
    \includegraphics[width = \columnwidth]{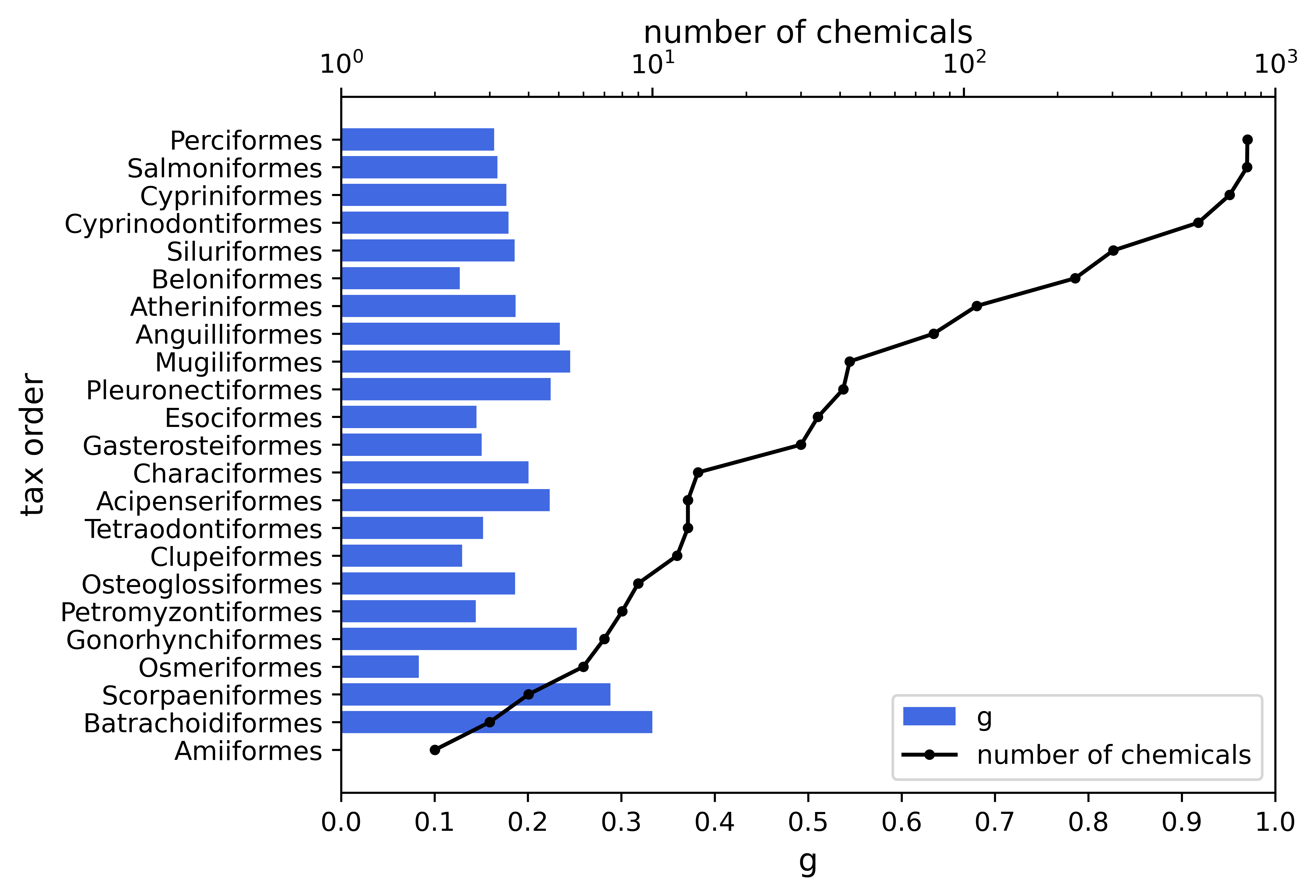}
    \caption{The bars depict, for each order, the \textit{taxonomic disagreement}, $g$, as defined in the main text. If $g=0$ for an order, all chemicals tested on that order give the same result also on the other orders. If $g>0$, that order reacts to chemicals differently from other orders.
    The black line (top $x$ axis) says how many chemicals were tested on that order.}
    \label{fig:g}
\end{figure}

In Fig.~\ref{fig:g} we show the mean value of $g$ for each taxonomic order (excluding those for which only a single experiment is available, see \ref{app:data}). We chose the order for representation clarity, but as we explained this can be done in a similar way with any taxonomic rank. 

The value of $g$ is positive and steady across the orders, indicating that experiments on one order are not fully representative of other orders, and that there is no single order that represents better or worse the other orders. The only exception would seem \textit{Amiiformes}, but this datum is poorly significant because we only have two experiments on this order (the black line in Fig.~\ref{fig:g} indicates the number of available experiments).

\subsection{The \texorpdfstring{\C}{C }setup: predictions based on chemical properties alone}\label{sec:tox}

Predicting toxicity outcomes based on chemical properties is the domain of quantitative structure-activity relationships (QSARs)~\cite{cherkasov:14}. 
Most often, existing QSARs are trained on the impact that a set of chemicals has on a single species. In our notation, this amounts to deducing $y_2$ from $\vec x_\mathrm{ch}$. 
Approaches of this kind are sometimes called \textit{global}, but since we refer to generic ways of using only $\vec x_\mathrm{ch}$,
we call this approach the \C setup.
In this section, we show the outcome of this kind of approach on our dataset.

There are two ways in which we can train models using only $\vec x_\mathrm{ch}$ (and ignoring $\vec x_\mathrm{tax}$ and $\vec x_\mathrm{ex}$):
\begin{enumerate}
    \item Train and test the models on the whole dataset, merging the $\vec x_\mathrm{ch}$ related to different taxa. %This reflects a situation in which one combines different data sources in order to obtain a larger dataset \mbj{[misleading!]}. 
    The performances of the different models are shown in Tab.~\ref{tab:acc-noneco-mixing}. 

    \item Train a model with data coming from a single species, and assess how well it generalizes to other species. 
    As training taxa we use Rainbow Trouts (\textit{Oncorhynchus mykiss}) and Fathead Minnow (\textit{Pimephales promelas}), which are commonly used~\cite{ankley:06} (we have $2500$ and $3000$ entries for each, see \ref{app:data}). 
    The performances of the different models are shown in Tab.~\ref{tab:acc-noneco-nomixing}.\footnote{
We do not report error estimates in Tab.~\ref{tab:acc-noneco-nomixing}, because restricting the dataset to a single species made the training set significantly scarcer, so decided not to make a validation set and used the test set to select the hyperparameters. This means that the values in Tab.~\ref{tab:acc-noneco-nomixing} are at risk of being overestimated, but it guarantees that they are not underestimated, which is our aim, as it will become clear in Sec.~\ref{sec:ecotox}.}
\end{enumerate}

\begin{table}[t]
\centering
\resizebox{\columnwidth}{!}{
\begin{tabular}{c||c|c|c|c}
Model    & Accuracy & Recall    & Specificity & F$_1$        \\ \hline \hline
LR       & 0.841(5) & 0.32(1) & 0.969(6)    & 0.44(2) \\
3-NN     & 0.841(7) & 0.46(2) & 0.935(7)    & 0.53(3) \\
RF       & 0.855(5) & 0.391(9)  & 0.969(8)    & 0.515(9)  \\
MLP      & 0.84(2)& 0.5(1)  & 0.93(1)   & \textbf{0.55(8)}    \\
S-RASAR  & 0.823(7) & 0.32(2) & 0.946(4)    & 0.42(2) \\
DF-RASAR & 0.848(8) & 0.30(3) & 0.983(6)    & 0.44(3)
\end{tabular}
}
\caption{Estimates using only the information on the chemical.
Metrics of different models, trained and tested on $\vec x_\mathrm{ch}$, \emph{i.e.} without distinguishing among taxa. 
We highlight the best performance in bold.
The numbers in parentheses are an uncertainty estimate (see Sec.~\ref{sec:errors}).
} 
\label{tab:acc-noneco-mixing}
\end{table}

\begin{table}[t]
\centering
\resizebox{\columnwidth}{!}{
\begin{tabular}{c|c||c|c|c|c}
Model    & Training Species   & Acc.  & Recall & Spec. & F$_1$    \\ \hline \hline
1-NN     & Fathead   Minnow & 0.805 & 0.668  & 0.880 & 0.709 \\
RF       & Fathead   Minnow & 0.814 & 0.647  & 0.906 & \textbf{0.712} \\
MLP      & Fathead   Minnow & 0.781 & 0.352 & 0.916 & 0.402\\
S-RASAR  & Fathead   Minnow & 0.807 & 0.572  & 0.936 & 0.678 \\
DF-RASAR & Fathead   Minnow & 0.817 & 0.606  & 0.933 & 0.702 \\\hline
1-NN     & Rainbow   Trout  & 0.820 & 0.827  & 0.816 & 0.753 \\
RF       & Rainbow   Trout  & 0.843 & 0.800  & 0.865 & \textbf{0.773} \\
MLP      & Rainbow   Trout  & 0.798 & 0.684 & 0.827 & 0.583 \\
S-RASAR  & Rainbow   Trout  & 0.837 & 0.767  & 0.872 & 0.758 \\
DF-RASAR & Rainbow   Trout  & 0.830 & 0.788  & 0.850 & 0.755 \\
\end{tabular}
}
\caption{Out-of-domain generalization, for different models, using only $\vec x_\mathrm{ch}$.
Here, models are trained on data coming from a single species (Rainbow Trout or Fathead Minnow),
and they are tested on all the other taxa contained in the data set.
}
\label{tab:acc-noneco-nomixing}
\end{table}

From Tabs.~\ref{tab:acc-noneco-mixing} and~\ref{tab:acc-noneco-nomixing}, it seems less advantageous to blindly mix all the taxa, than to train on a single species, and then na\"ively extrapolate. The reason for this can be that the former method involves merging data coming from different taxa \emph{before training}. This implies that i) the dataset becomes smaller and unbalanced (see \ref{app:data} - anyhow, this can be partly offset by using the F1-score as a metric), and ii) that the labels $y_2$ are perturbed {in an unknown way}, since a label that correctly applies to one taxon does not necessarily apply to another. Further tests of the \C approach are shown in \ref{app:stratified}.

\subsection{The \texorpdfstring{\CTE}{CTE }setup: combining chemical, taxon and experimental details}
\label{sec:ecotox}
\begin{figure}
    \centering
    \includegraphics[width=\columnwidth]{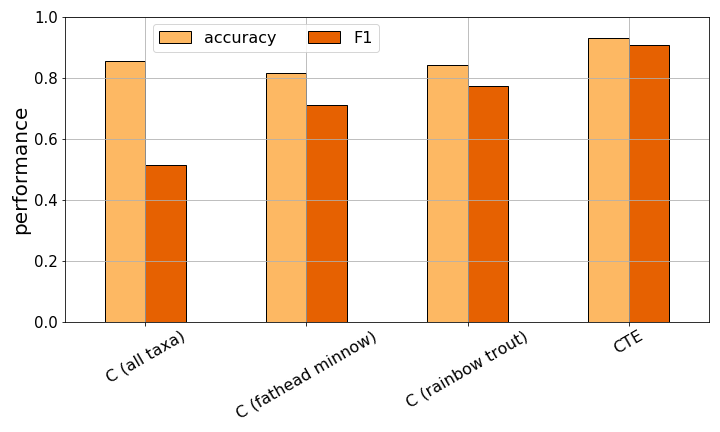}
    \caption{Accuracy (red) and F1 score (green) of random forests trained with the \C setup (first three columns, data corresponding to Tabs.~\ref{tab:acc-noneco-mixing},\ref{tab:acc-noneco-nomixing}), and with the \CTE setup (fourth column, Tab.~\ref{tab:acc-eco}).
    % The hatches on the bars highlight the kind of data source (or equivalently, whether the data comes from Tab.~\ref{tab:acc-noneco-mixing},~\ref{tab:acc-noneco-nomixing} or~\ref{tab:acc-eco}).
    }
    \label{fig:accf1}
\end{figure}

We now turn to what we define the \CTE setup, that combines all the different sources of information. At this stage, we treat $\vec x_\mathrm{ch}$, $\vec x_\mathrm{tax}$ and $\vec x_\mathrm{ex}$ on the same footing. For KNN and RASAR we use the distance defined in Eq.~\eqref{eq:naif-dist} (details on model training can be found in \ref{app:models}).
In Tab.~\ref{tab:acc-eco}, we show the performance of each single model, and on Fig~\ref{fig:accf1} we plot the performance of RFs using the \C versus the \CTE approaches. 
\begin{table}[t]
\centering
\resizebox{\columnwidth}{!}{
\begin{tabular}{c||c|c|c|c}
Model         & Accuracy & Recall   & Specificity & F$_1$       \\ \hline \hline
LR            & 0.882(2) & 0.834(5) & 0.910(2)    & 0.841(4) \\
3-NN          & 0.918(1) & 0.888(5) & 0.892(3)    & 0.890(2) \\
RF            & \textbf{0.932(1)} & \textbf{0.903(3)} & 0.949(3)    & \textbf{0.909(2)} \\
MLP           & 0.913(4) & 0.887(4) & 0.929(6)    & 0.887(5) \\
S-RASAR       & 0.910(1) & 0.858(7) & 0.941(5)    & 0.877(2) \\
DF-RASAR      & 0.919(2) & 0.861(4) & \textbf{0.953(2)}    & 0.888(3)
\end{tabular}
}
\caption{Estimates of $y_2$ from the whole dataset $\vec x= (\vec x_\mathrm{ch}, \vec x_\mathrm{tax}, \vec x_\mathrm{ex})$, which combines information on chemical, taxonomy and experimental conditions.
}
\label{tab:acc-eco}
\end{table}
The better performance of models trained with a \CTE approach stands out. Even a simple linear model such as the LR reaches good performances ($\text{F}_1\simeq0.82$) outperforming all the models obtained with both \C approaches (see Tabs.~\ref{tab:acc-noneco-mixing} and~\ref{tab:acc-noneco-nomixing}). Moreover, the gap between the LR and our best non-linear models is below 10\%, demonstrating that much of the information can be captured through simple linear relationships.
The best model is the RF, followed by the DF RASAR.\footnote{
 In \ref{app:stratified} we provide further comparisons between the \C and \CTE approaches, showing that with different kinds of train/test splitting the \CTE approach still outperforms the \C, and that RF and DF-RASAR perform best.
}
The better performance, lower data requirement, and higher simplicity of implementation are strong factors in favor of the RF. However, we must keep in mind that the DF RASAR is explicitly designed for problems with a high availability of data from different biological effects, while in our dataset the effect ``mortality" is dominant with respect to the other effects, which the DF RASAR uses as input to predict mortality (see \ref{app:data}, and \ref{app:models}).
Furthermore, there is space for improving of RASAR models. One way is to replace the Euclidean distance (Eq.~\eqref{eq:naif-dist}) used by the RASAR with a distance that gives a different weight to different sources of information (Eq.~\eqref{eq:dist}), as described in Sec.~\ref{sec:def-alphas}. 
In Tab.~\ref{tab:alphas} we show the performances obtained by using this procedure.
\begin{table}[t]
\centering
\resizebox{\columnwidth}{!}{
\begin{tabular}{c||c|c|c|c}
Model      & Accuracy          & Recall            & Specificity       & F$_1$                \\ \hline \hline
1-NN       & 0.918(2)          & \textbf{0.898(3)}          & 0.931(4)          & 0.892(3)          \\
3-NN       & 0.917(2)          & 0.896(2)          & 0.930(4)          & 0.890(2)          \\
5-NN       & 0.908(4)          & 0.875(6)          & 0.928(3)          & 0.877(4)          \\
S-RASAR    & 0.914(1)          & 0.896(6)          & 0.924(5)          & 0.886(1)          \\
DF-RASAR   & \textbf{0.923(3)} & 0.884(5)          & \textbf{0.946(3)} & \textbf{0.908(6)}
\end{tabular}
}
\caption{Estimates of $y_2$ from the whole dataset $\vec x= (\vec x_\mathrm{ch}, \vec x_\mathrm{tax}, \vec x_\mathrm{ex})$, using the distances defined in Eq.~\eqref{eq:dist}. We only list the models that rely on the definition of the distance [Eqs.~\eqref{eq:naif-dist} and~\eqref{eq:dist}], since the others are unaffected.}
\label{tab:alphas}
\end{table}
We note an improvement throughout most of the metrics of all the models (K-NN, S-RASAR and DF-RASAR), especially the DF-RASAR, which reaches $a=92.3\%$ and F$_1=90.8\%$.\footnote{Note that, even though, for completeness, we provide the KNN results for K=1,3,5, comparisons between tables \ref{tab:acc-eco} and \ref{tab:alphas} should compare KNN as a single algorithm.
The specificity of S-RASAR shows a decrease but is compensated by the higher recall score, achieving a higher F1-score, which is generally the most desirable result.}
% Since for the RASAR we used the $\alpha$s obtained from the 1-NN, we anticipate that tuning them on the RASAR can give even better performances~\cite{wu:21}.

From Sec.~\ref{sec:repeated-exp} we know that the maximum achievable accuracy is not 100\%, but rather an unknown value $\tilde{a}$. We cannot measure $\tilde{a}$ directly, but we can estimate it by measuring $A_\mathrm{up}$ [Eq.~\eqref{eq:aup}]. From Table \ref{tab:exp-acc}, we see that the accuracy of the experiments is at most $A_\mathrm{up}=0.958$, so we can use this value to rescale our performances.
For example, the rescaled performance of the RF becomes $0.932/0.958\simeq0.973\approx97\%$. 
On another side, we note that most of our models perform better than what would be indicated through the conditional estimator of $\tilde{a}$ (Table \ref{tab:exp-acc}), suggesting that this metric is a too pessimistic estimate of $\tilde{a}$.

\subsection{Feature Importance}

\subsubsection{Importance from Random Forests}
We now discuss the relative importance of the features, when using our best model (the RF of Tab.~\ref{tab:acc-eco}). These importances are estimates of how relevant each feature is for the predictions.
Since there is no single best method to calculate feature importances, we evaluate them through two most commonly used methods: permutation-based and impurity-based (using Scikit-Learn libraries~\cite{scikit-learn}). 
While the permutation-based importance falls short with collinear features, the impurity-based tends to overestimate the relevance of high-cardinality categorical features~\cite{altmann:10}.
Even though we do observe some (expected) differences between the two methods, the conclusions we draw are consistent across both kinds of analysis. We here show the results with the impurity-based method, and provide the permutation importance analysis in \ref{app:importance}.
 
\begin{figure}[tb]
    \centering
    \includegraphics[width=\columnwidth]{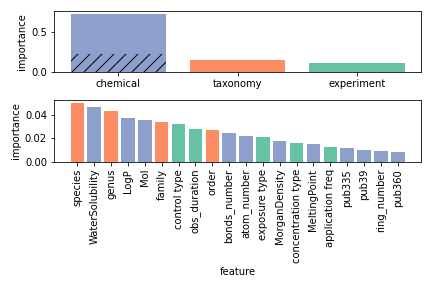}
    \caption{\textbf{Top}: Feature importance from the RF. Every bar corresponds to a collection of features. The bottom (hatched) part of the bar corresponding to the chemicals is related to the chemical fingerprints, whereas the top part depicts the importance of the pubchems. \textbf{Bottom}: feature importance of the single features. For visualization purposes, we only include the 20 most important features (see \ref{app:data} for a description).}
    \label{fig:feature_importance_RF}
\end{figure}

In the top chart of Fig.~\ref{fig:feature_importance_RF}, we coarsen the feature importances in three groups, corresponding to $\vec x_\mathrm{ch}, \vec x_\mathrm{tax}$ and $\vec x_\mathrm{exp}$. It can be clearly seen that $\vec x_\mathrm{tax}$ and $\vec x_\mathrm{exp}$ contribute to almost 30\% of the information.
The contribution of the features related to taxonomy and experimental details is even more striking when we look at them one by one:
in the bottom set of Fig.~\ref{fig:feature_importance_RF}, we show the importance of the single 20 most important features. 
The single most important feature appears to be the species. The most important features from $\vec x_\mathrm{ch}$ are water solubility and LogP. The control type and observation duration (details in \ref{app:data}) appear to be the most influential experimental details, and the most important pubchems have the labels 335, which indicates the presence of the substructure \texttt{C($\sim$C)($\sim$C)($\sim$C)($\sim$H)}, and 39, which indicates the presence of more than 4 \texttt{Cl} atoms~\cite{pubchem-id}. Also with the permutation importance we find that the taxonomy- and experiment-related features, taken one by one, carry a large importance compared with the chemical descriptors taken one by one.
These observations are consistent with Ref.~\cite{tuulaikhuu:17}, where the species is identified as the most important taxonomic descriptor, and the LogP as the most important chemical. %However, collinearity must be taken into account when interpreting Fig.~\ref{fig:feature_importance_RF}. For example, the importance of the fish family would likely increase if the species was not used. This means that we expect that lower taxonomic ranks have a higher importance, but that the lower importance of higher taxonomic ranks does not make them irrelevant, as it was previously alluded~\cite{tuulaikhuu:17}. A similar reasoning also applies \mbj{to} chemical descriptors.

\subsubsection{Importance from \texorpdfstring{$\alpha$}{a}}

Since the distances used in Eq.~\eqref{eq:dist} are normalized to one, the optimal values of each $\alpha$ tell us how relevant each set of features is.
In particular, $\alpha_\mathrm{cat}$ is an indicator of how relevant the data on experiment and taxon, $\vec x_\mathrm{ex}$ and $\vec x_\mathrm{tax}$ are with respect to the information on the chemical.
In Tab.~\ref{tab:alpha-values} we show the optimal values of $\alpha$ for several models. The first thing that we point out is that $\alpha_\mathrm{cat}$ is not zero, meaning that taxonomy and experimental details provide useful information.
This is consistent with our previous feature importance analysis, and the remark that $\alpha_\mathrm{cat}$ is $\sim20$ times smaller than the other two descriptors suggests that these models rely more heavily on $\vec x_\mathrm{ch}$ than the RFs. 
Second, we notice that now the pubchems ($\vec x_\mathrm{pubc}$) seem to often play a smaller role than the explicit chemical properties ($\vec x_\mathrm{pr}$), than what we had seen with the RFs. We see two possible reasons for this: (i) the same information is contained both in the pubchems and in the properties, and which of the two is used is decided by the model while training. This is expected at least to a degree, since physicochemical properties can be predicted from the molecular structure~\cite{mansouri:18}. (ii) The distances in the pubchem space do not discern which chemical properties are different, and assign the same weight to all kinds of variation.
This is also true for the chemical properties, $\vec x_\mathrm{pr}$, but the properties are fewer, not binary, features. One way to come around this would be to treat different groups of pubchems components separately, by introducing more $\alpha$ constants.

\begin{table}[t]
\centering
\begin{tabular}{c||ccc}
Model & $\alpha_\mathrm{Euc}$    & $\alpha_\mathrm{pubc}$ & $\alpha_\mathrm{cat}$ \\ \hline \hline
1-NN     & 1 &  0.5623    & 0.0749 \\
3-NN     & 1 &  1.7433    & 0.2212 \\
5-NN     & 1 &  0.9459    & 0.0833 \\
S-RASAR  & 1 &  0.5623    & 0.0749  \\
DF-RASAR & 1 &  0.5623    & 0.0749
\end{tabular}
\caption{Values of $\alpha_\mathrm{Euc}$, $\alpha_\mathrm{pubc}$ and $\alpha_\mathrm{cat}$ [Eq.~\eqref{eq:dist}] for different models. 
These parameters are found during CV, and indicate the relative importance of the three sources of information (chemical fingerprints, pubchems, taxonomy \& experimental details) for some of our models (see main text). 
For the RASAR models we use the ones of the 1-NN, since it is both the best K-NN model, and the one that is conceptually most similar to the RASAR.}
\label{tab:alpha-values}
\end{table}

\subsection{Multiclass models}
The whole analysis that we showed for binary classification was repeated in the case of 5-class labeling. Other than requiring to adjust the RASAR in order to also include an arbitrary number of classes (\ref{app:models}), the analysis was analogous to the binary case. Also the results are similar, so we relegate them to \ref{app:multiclass}. Here, we only make some remarks:
\begin{itemize}
    \item The performances of our models are lower for multiclass, but also the self consistency of the data is lower, since with more classes it is more likely that the same experiment repeated twice falls into two different categories. If we normalize the recall by the upper bounds shown in Tab.~\ref{tab:exp-acc}, we obtain a rescaled accuracy of $89\%$, and a rescaled recall of $95\%$.
    \item The performance of the multiclass DF-RASAR model matches that of the RF.
\end{itemize}

\section{Conclusions}
In this paper, we addressed the problem of predicting the acute toxicity of chemicals on different taxa through machine learning. We focused on fish datasets where one is interested in knowing about the effect of many chemicals on a large number of taxa, and the available data only includes few combinations of taxon and chemical. We addressed both binary and 5-class classification, with similar results.

\paragraph*{Gain from including taxonomic and experiment information}
We showed that taking into account the information on the taxon and on the experimental conditions is highly beneficial, providing over a 10\% gain in F1 score (and other metrics, such as accuracy) with respect to the standard procedures of dealing with the same dataset. Remarkably, a linear model trained on a dataset which includes taxonomic information performs better than more sophisticated models which only make use of the chemical information.
Additionally, we showed how one can take into account the peculiarities of each source of data to obtain a further gain.

\paragraph*{Best models and comparison to RASARs}
The best performances are obtained by the RF, followed by the DF-RASAR and other models that obtained performances that were not too far. This similarity between different models was already found previously~\cite{li:17b,hou:20}, and the superiority of RFs had already been pointed out~\cite{hou:20}.
Even though the DF-RASAR obtains good performances, they do not seem to outperform other traditional machine learning models. In particular, the S-RASAR does not outperform K-NN, making its deployment less justifiable. The DF-RASAR, instead, has several advantages in the integration of the data from different effects, and a potential for improvement through some small variations.

\paragraph*{Best performance in terms of maximum achievable performance}
For binary classification, we obtain a maximum prediction accuracy of 93.2\%. However, since this number comes from a test set which contains some inconsistent examples, this number is effectively higher than it appears, since it should be normalized by the accuracy of the test set. We cannot estimate it, but we see that it is roughly upper-bounded by $A_\mathrm{up}=0.958$. By rescaling appropriately through $A_\mathrm{up}$, we can argue that the real accuracy is around $97\%$. Similar reamarks are valid for the multi-class setting.

\paragraph*{Metrics for comparing ML models with experiment reproducibility}
If we use the metrics presented in Ref.~\cite{luechtefeld:18}, \emph{all} our models (as long as we use a \CTE approach), outperform animal experiments. As also mentioned in Ref.~\cite{alves:19}, this statement should thus be polished, given the non-rigorous nature of this kind of estimators. To deal with this, instead of trying to estimate the accuracy (as well as other metrics) of the experiments, we estimated a heuristic upper bound. With these upper bounds, we still find evidence that our best models (RASAR, RF, KNN and MLP) are outperforming animal tests (Sec.~\ref{sec:repeated-exp}). However, these comparisons should be taken with a grain of salt.
In fact, the main problem is that, as stated earlier, this kind of comparisons is not rigorously defined. When we estimate the accuracy of the experiments, we have no truth value with which we can compare; we only have the results of repetitions of the same experiment. Therefore, we cannot estimate it without making strong assumptions. When we then calculate the accuracy of our models, we test them with a random selection of outcomes from experiments. In other words, we are now making the implicit (wrong) assumption that the data in the test set is not mislabeled (\textit{i.e.} we are assuming that all the experimental outcomes faithfully reflect the ground truth). To overcome this issue, we made heuristic arguments on how the test accuracy of the models would be affected by the noise in the labels. However, these arguments are also not fully representative, since the accuracy of the labels is calculated on single repetitions of the same experiments (and assuming that all experiments fluctuate in the same way), but the data used for training and testing coarsens repeated entries in a single one, since we take the median LC50. Taking the median is desirable, since it arguably produces higher-quality data, but makes it even more far fetched to compare the performance of the models with that of the experiments.
Anyhow, albeit not mathematically rigorous, these metrics arguably give a reasonable indication of the reliability of experimental results, and can be used to get a gross understanding of how accurate machine learning models are.
In our case, they indicate that some of our models are close to the maximum obtainable performance, but only as long as we include taxonomic variability as an input variable.
This means that, for the data we analyzed, the model selection is only marginally relevant with respect to the right choice and curation of the input features.

\paragraph*{Most important descriptors}
We discriminated which input features played the most important role. Most of the important features belong to the chemical, demonstrating why it is still possible to obtain decent predictions based on the chemical only. Nonetheless, the role of taxonomy and experimental details is relevant (\textit{e.g.} it explains 30\% of the inter-trees variability in the RFs), and the single most important feature is the species (the species descriptor is more crucial than the LogP towards assessing toxicity), which is in agreement with previous observations~\cite{tuulaikhuu:17}.
Moreover, in this work we restricted our analysis to fishes, but expanding the focus to wider taxonomic ranges would certainly further increase the importance of including the taxonomy.

\paragraph*{Relation to human toxicology}
Training models that are efficiently able to generalize across taxa is central in ecotoxicology, but it is also crucial for human-centered toxicology. In fact, the investigation of toxic effects on humans passes through the experimentation on mammals, such as rats, mice, guinea pigs and rabbits, and there is no guarantee that these species react in the same way as humans to novel toxicants.
In recent years \textit{in-silico} methods, including machine learning, have been used to predict human toxicity - most often based on oral toxicity tested on rodents-~\cite{bhhatarai:15,gadaleta:19,nedelcheva:19,chushak:21,mansouri:21}, under the implicit assumption that humans react in the same way as the tested mammals.
Therefore, mastering generalization across taxa can help data-driven approaches to infer better from animals to humans.

\paragraph*{Replacing \vivo animal testing}
In the long term, it would be desirable to replace \vivo testing through machine learning. How far are we from that? 
In order to do this, we would like \textit{in-silico} predictions to be as accurate, or almost, as directly performing an \vivo test.
Our results represent a fraction of all the possible chemicals and taxa that one could be interested in testing. In order to expand the domain of applicability of our models, we would need to use a wider range of chemicals and taxa both for training and, more crucially, for testing. A known problem in machine learning is that the performances obtained in the test set often degrade when the models are deployed in real application settings, because the datasets used to test these models did not comprehensively represent the whole feature space. This problem is also known as dataset shift~\cite{quinonero:09,morenotorres:12}. One main source of dataset shift is the fact that the sampling of data points is not completely random, but follows patterns. This is certainly the case in (eco)toxicology, where taxa and chemicals that appear in the databases are chosen according to specific needs. 
Therefore, before the deployment of machine learning for risk assessment can become an option, it is necessary to have extensive-enough test sets, and a thorough study of dataset shift. Even then, it is foreseeable that for exotic chemical/taxonomic categories the machine learning predictions might need to be confirmed through tests.
Therefore, it is unlikely that machine learning alone will soon be able to completely replace animal testing, but it can certainly allow to reduce it drastically.
In addition, given that it is anyhow hopeless to test all chemicals on all taxa, machine learning can be effectively used out of the box, even at the current stage of development, for prioritization: the results of \textit{in-silico} methods can guide us on which chemicals/taxa should be tested first.

\section*{Acknowledgement}
We thank C. Sch\"ur and L. Gasser for detailed feedback on the manuscript, M. Leone, G. Macchi and M. Vicentini for their preliminary results at the start of this project. 
We are grateful to L. Gasser for sharing the base version of the code for the stratified splitting of the data, and for extra feedback on model training.
We thank F. Balk, F. Hartig, M. Mair, F. P\'erez-Cruz, P. Reichert, A. Scheidegger, N. Schuwirth and A. Zupanic for inspiring conversations.

\section*{Author Information}
Jimeng Wu and Simone D'Ambrosi share first authorship.

\section*{Funding Sources}
This project was funded by the SDSC grant ``Enhancing Toxicological Testing through Machine Learning" (project N$^o$ C20-04).

\section*{Data and code availability statement}
Our analyses were performed on an openly available data set~\cite{ecotox}.
Our code for data cleaning and for model training and testing is freely available at \href{https://github.com/mbaityje/ML-Tox}{\url{github.com/mbaityje/ML-Tox}}.

% FROM HERE IT IS APPENDIX

\appendix

\section{Data} \label{app:data}

We use the Ecotox database, provided by the Environmental Protection Agency of the USA~\cite{ecotox}, and download the whole ASCII database. 
The dataset contains information on the taxonomy of the tested organism, on the conditions under which the experiment was brought through, a CAS identifier of the used chemical, and several endpoints related to the studied effects. %\wjm{We cleaned this dataset and created two data tables. One on all experiments on mortality and the other containing all other experiments. The former was used in all models in this paper, and the latter was utilized only in the datafusion model.}

% \wjm{The selected features are as follows: 
% \begin{itemize}
%     \item Chemical: "ring number", "triple Bond", "double Bond", "alone atom number", "oh count", "atom number", "bonds number", "molecular weight", "Morgan Density", "LogP", "water solubility", "melting point", "pubchem2d";
%     \item Taxonomic: "class", "tax order", "family", "genus", "species";
%     \item Experimental conditions: "observation duration", "control type", "concentration type", "exposure type", "application frequency", "media type".
% \end{itemize}}

\paragraph{Chemicals}
From the CAS codes provided in each EcoTox entry, we extract the SMILES string representing the chemical. 
We use the rdkit python library~\cite{rdkit} to extract the following features from the SMILES: number of atoms, number of alone atoms, number of OH groups, number of bonds, number of double bonds, number of triple bonds, and number of rings, LogP (octanol-water partition coefficient) and Morgan Density.
Additionally, we extract molecular weight, water solubility and melting point from the Comptox database~\cite{williams:17}.
From the SMILES we also obtain the Pubchem2D vector through the pubchempy library~\cite{pubchempy}.
In a small number of cases we are not able to retrieve crucial information such as the SMILES or the Pubchem vectors. We decide to drop those data entries.

\begin{figure}[tb]
    \centering
    \includegraphics[width = \columnwidth]{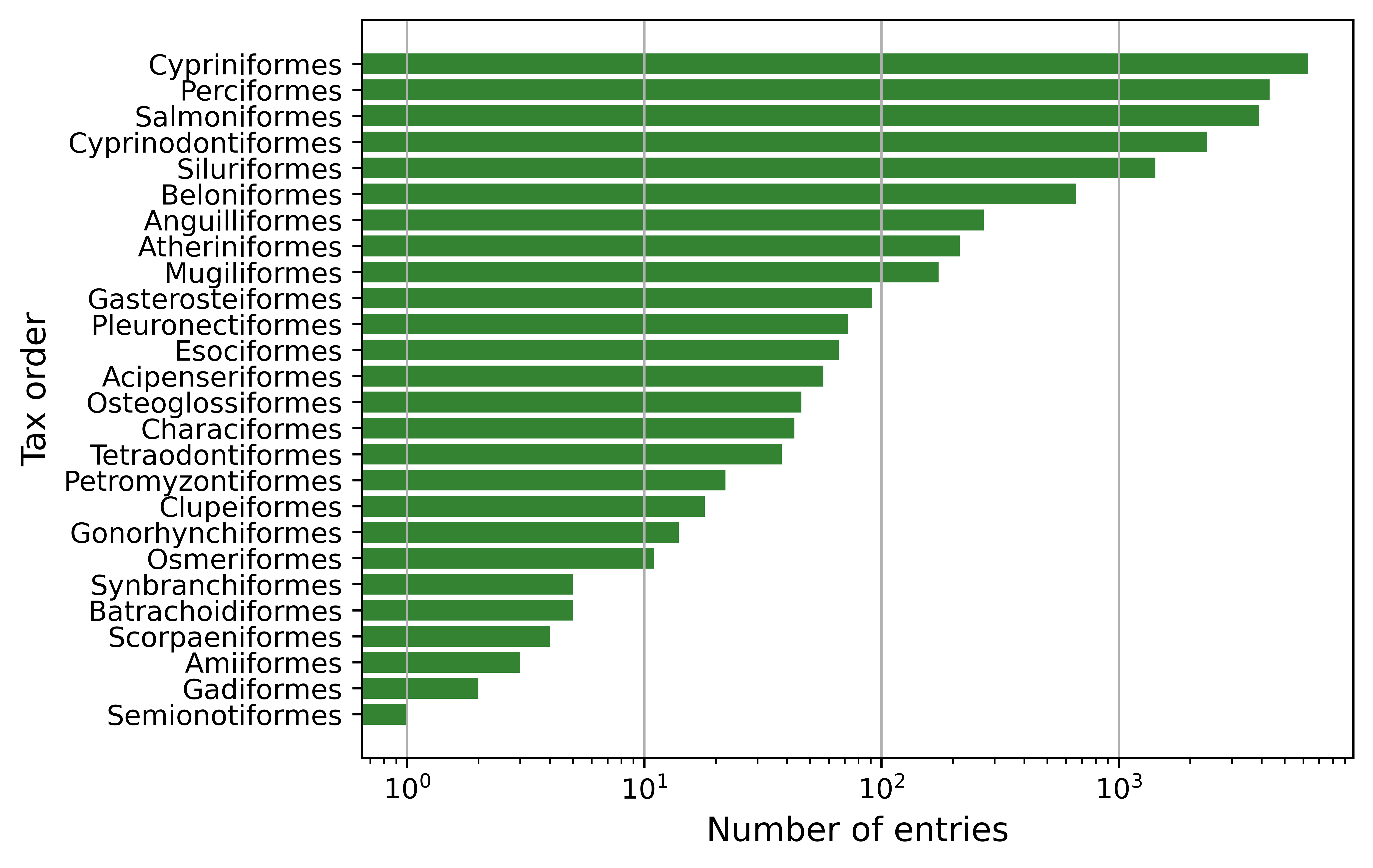}
    \caption{Histogram of the different orders appearing in our dataset after data preparation.}
    \label{fig:histo_taxa}
\end{figure}

\begin{figure}
    \centering
    \includegraphics[width=\columnwidth]{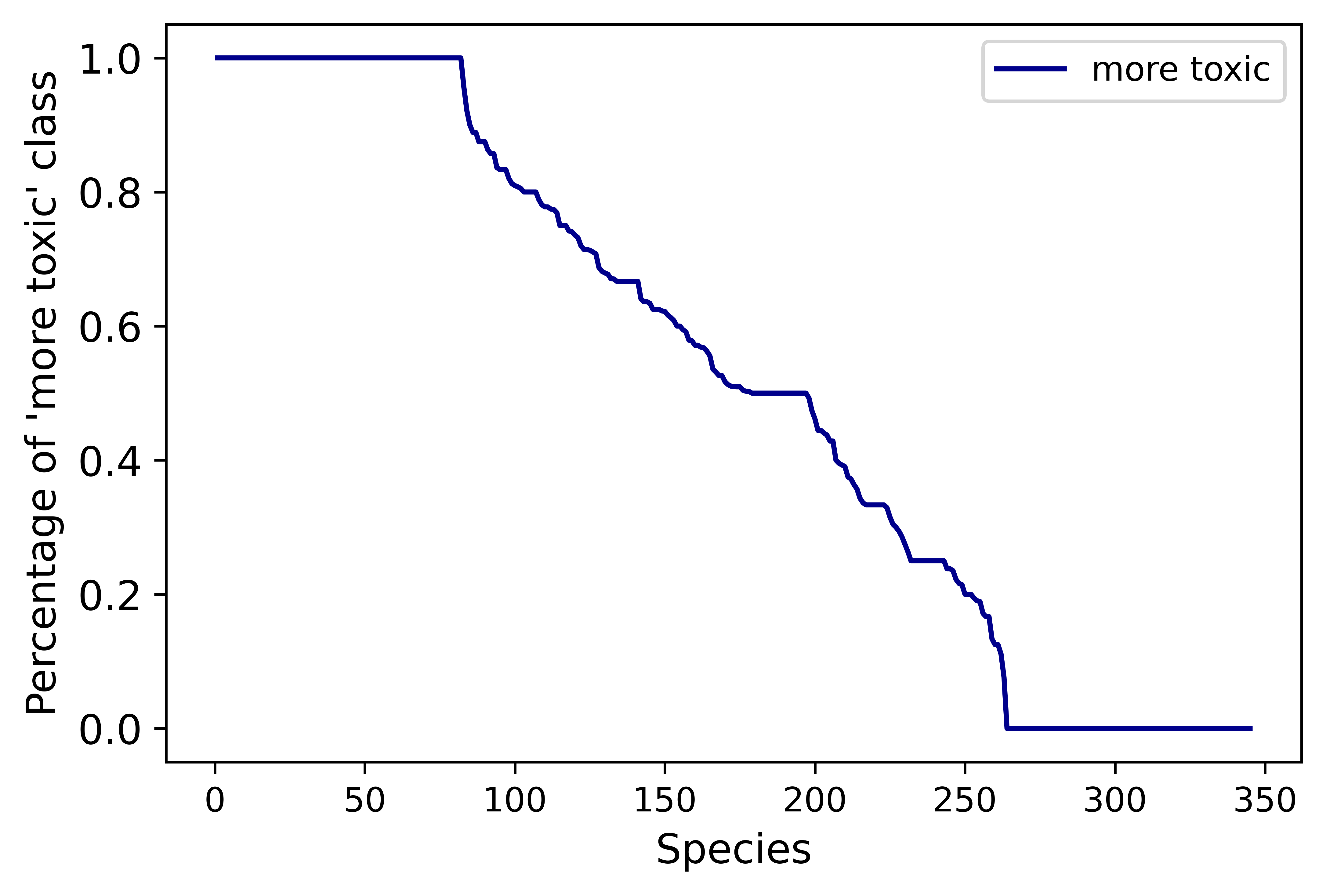}
    \caption{For each species, we report the fraction of chemicals that results \emph{more toxic} to it. Species on the $x$ axis are ordered by rank, in order to have a non-increasing curve.}
    \label{fig:Toxic_distribution_species}
\end{figure}

\paragraph{Taxonomy}
We filter the dataset keeping only the experiments on fishes. This implies that we can discard the high-rank taxonomic features, ending up with 2 classes (Actinopterygii and Cephalaspidomorphi), 26 orders, 83 families, 219 genuses and 345 species. We also exclude organisms at the embryo life stage.
Note that the data is not equally distributed among taxonomic categories. For example, in {Fig.~\ref{fig:histo_taxa}} we show how many times every order appears in our dataset (after cleaning), which is highly imbalanced. 
We also show, in Figure~\ref{fig:Toxic_distribution_species}, the fraction of chemicals that result \emph{more toxic} to it.

% 'measurement', 'conc1_type', 'conc1_mean',
%       'conc1_unit', 
\paragraph{Experimental conditions}
As for the experiment conditions, we keep those that have a sufficiently small number of missing entries:
'observation duration' (24h, 48h, 72h, 96h - most are 96h),
'control type' (Concurrent control, Insufficient, Unknown, Satisfactory, Carrier or solvent, Multiple types, Unsatisfactory, Baseline, Positive controls), 
'concentration type' (Active ingredient,
Dissolved,
Formulation,
Labile [free metal ion],
Not applicable,
Not coded, Not reported,
Total,
Unionized), 
'exposure type' (Static, Flow-through, Aquatic not reported, Renewal, Lentic, Pulse, Intraperitoneal, Subcutaneous, Spray),
'application frequency',
'media type' (fresh water, salt water). 
The names are self-explanatory, and we refer to the Ecotox documentation for a detailed description of the features (\href{https://cfpub.epa.gov/ecotox/help.cfm?sub=term-appendix}{\url{https://cfpub.epa.gov/ecotox/help.cfm?sub=term-appendix}}).

\begin{figure}[t]
    \centering
    \includegraphics[width=\columnwidth]{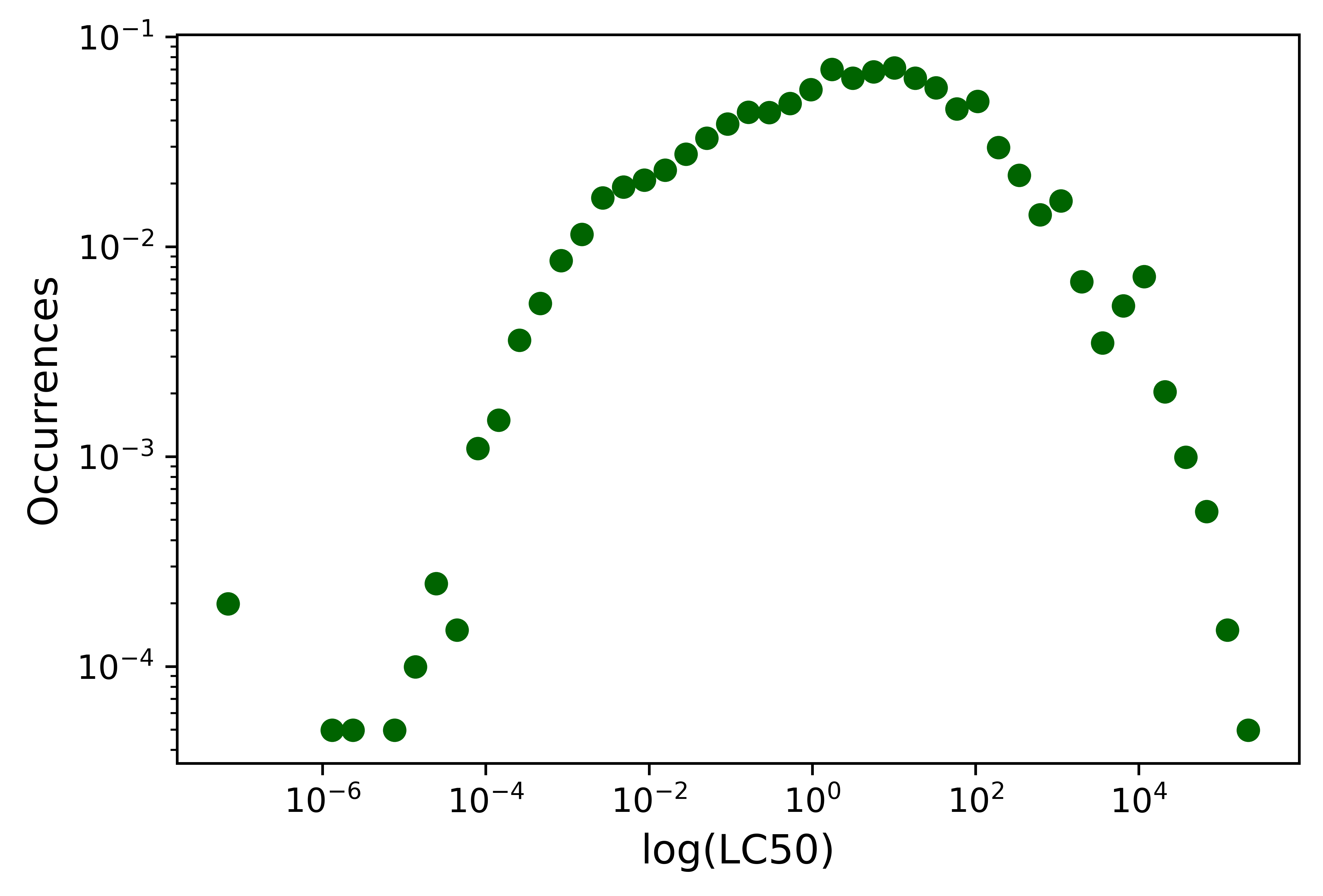}
    \caption{Distribution of the LC50 values in our dataset.}
    \label{fig:LC50_hist}
\end{figure}

\paragraph{Target}
We then filter on the desired endpoint and effect. For mortality experiments, the endpoints EC50 and LC50 are equivalent, so we keep both. The distribution of the LC50 in our dataset is shown in Fig.~\ref{fig:LC50_hist}.
For the other effects, that are needed by the DF RASAR as input, we take the EC50.

\paragraph{Imputation and aggregation}
We drop the lines with one or more missing features, and aggregate the experiments with the same CAS, taxonomy, and experiment conditions onto a single record, whose LC50 (or equivalent target) is the median LC50 from those repeated experiments. We then use the median LC50 to calculate the labels $y_2$ and $y_5$. While for the mortality data, which constitutes our target, we use the thresholds described in Eq.~\eqref{eq:binary} (or Eq.~\eqref{eq:multiclass} for multiclass), for the other effects, which are only used as input features for the DF-RASAR models, we use the median (or $20^\mathrm{th},40^\mathrm{th},60^\mathrm{th}$ and $80^\mathrm{th}$ percentiles for multiclass) in order to obtain maximally well-separated input features.

\paragraph{Standardization}
All the features are rescaled between 0 and 1, and with a prior logarithmic transformation when the quantity spans several orders of magnitude.
The scripts we used for the data cleaning are {freely} available~\cite{mltox-code}.

We end up with 20128 entries in our mortality dataset, including
$n_\mathrm{chem}=2199$ chemicals and $n_\mathrm{spec}=345$ species. 

The other studied effects (only for the DF-RASAR, see \ref{app:models}) are genetics (777 entries), enzyme (691 entries), biochemistry (684 entries), accumulation (608 entries), behavior (422 entries), physiology (242 entries), cells (139 entries), intoxication (98 entries), and multiple (52 entries).

\paragraph*{Class balance}

\begin{figure}[tb]
    \centering
    \begin{tabular}{@{}cc@{}}
    \includegraphics[width=0.45\linewidth]{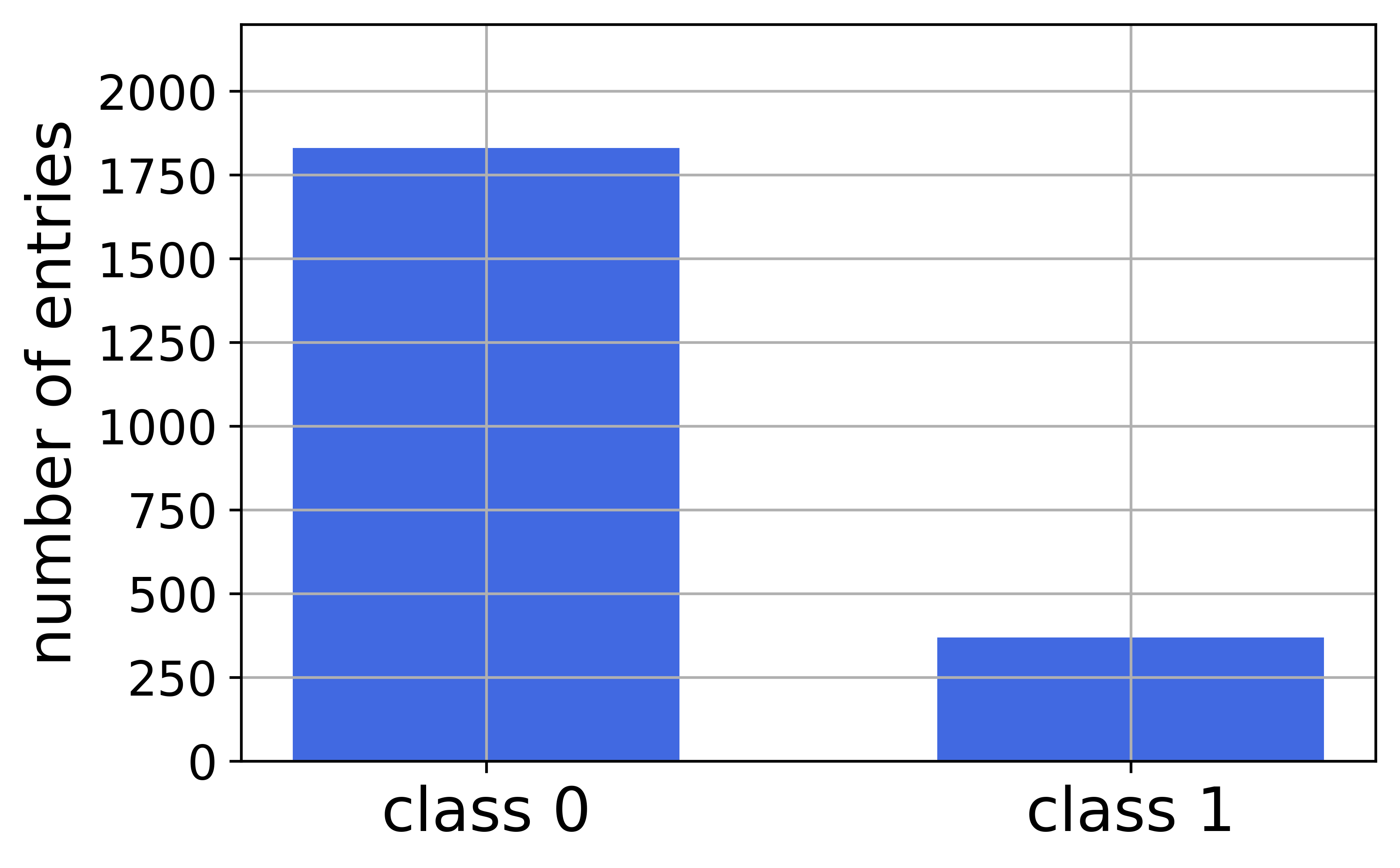}
    \includegraphics[width=0.45\linewidth]{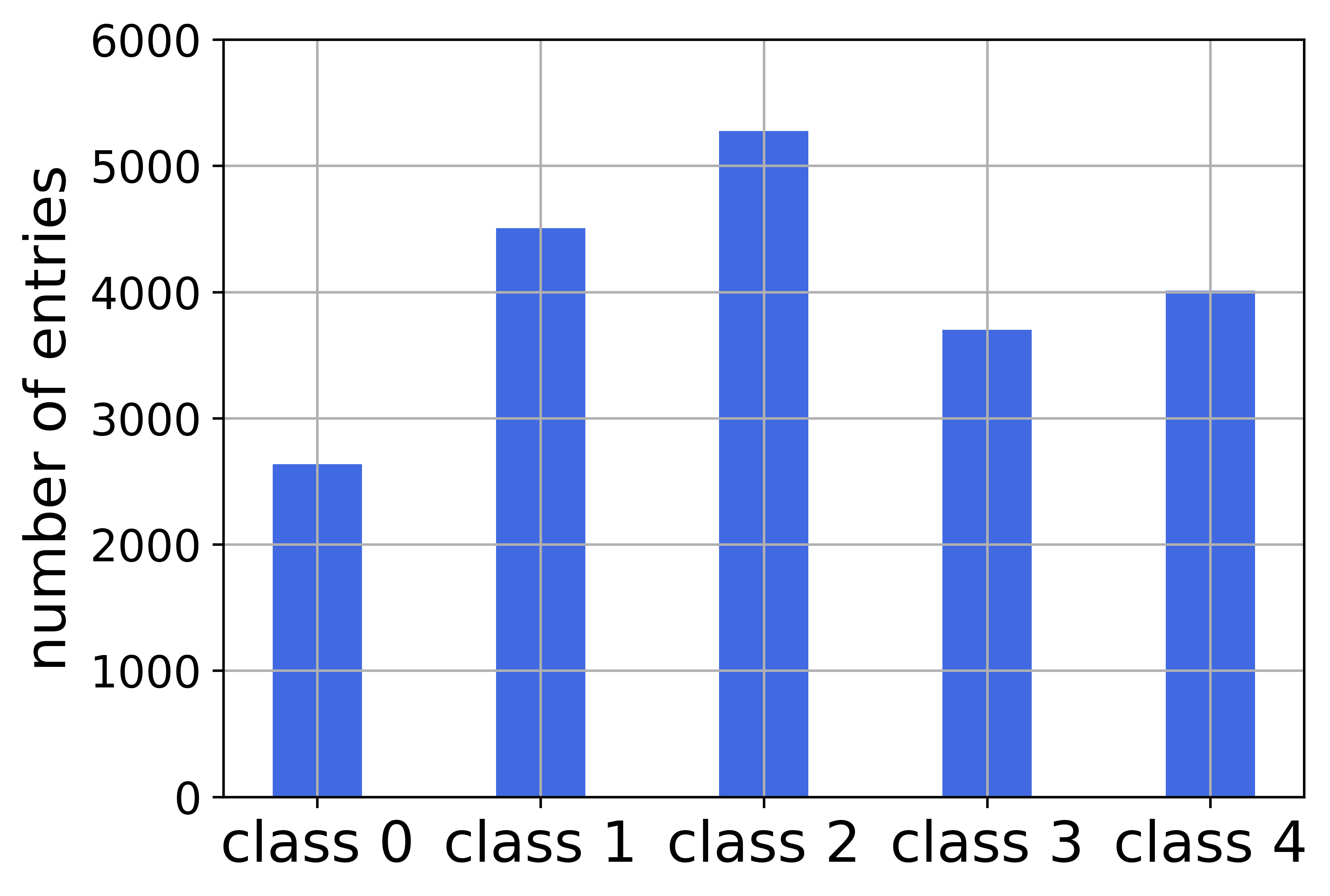}
    \end{tabular}
    \caption[class proportions]{Number of experiments per label after binary and 5-class labeling of the concentrations with only chemical features. The thresholds that define the classes are summarized in the appendix (Tab.~\ref{tab:classes}).}
    \label{fig:class_c}
\end{figure}

As shown in Fig.~\ref{fig:class}, the data set is balanced both in the binary and in the 5-class classification.
If we neglect all the features different from $\vec x_\mathrm{ch}$, we need to merge into a single entry all the lines related to different taxa/conditions and the same chemical, which makes the dataset less balanced. We end up having 461 positives and 1738 negatives and 543 non toxic, 640 hazardous, 555 toxic, 302 very toxic and 159 extremely toxic for multiclass (Fig.~\ref{fig:class_c}).
If we restrict to fathead minnows, we have 2830 entries.
If we restrict to rainbow trouts, we have 2508 entries.

\section{Models} \label{app:models}
In this section we explain, where necessary, the implementations of our machine learning models. All our models are trained by splitting the dataset (80:20) into a {training}
%train 
and a {hold-out} test set. The training set is used for 5-fold CV, implying a further 80:20 splitting into training and validation sets. The test set is not used until the very end, to produce the performance metrics.

Except for the RASARs, the models we use are quite standard, so we will not get into the details of most models, but rather in hyperparameter tuning and choices.

\subsection{Logistic Regression and Random Forest} \label{app:LR+RF}
For LR and RF we use the standard implementation from the Scikit Learn python library~\cite{scikit-learn,mltox-code}.
We rescale all numerical features through min-max normalization and perform one-hot encoding (using the OneHotEncoder function from Scikit Learn~\cite{scikit-learn}) on all categorical features in the preprocessing step. The hyperparameters adjusted in logistic regression include (the range of values in parentheses): $C$ ($10^{-3}$ - $10^{3}$), $l1$ ratio (0-1), and max iter (100-800).
In the random forest model, the adjusted hyperparameters are $n$ estimators (200-1000), maximum depth (10-20), minimum sample segmentation (2-10), and minimum sample leaf (1-32).

\subsection{Multi-Layer Perceptron}
As a first preprocessing step, we one-hot encode all the categorical features. Then, we apply a min-max scaler to the data so that the scaled data has a minimum of 0 and a maximum of 1. The lower bound of 0 is chosen to keep the sparsity of the categorical features after one-hot encoding.

We use a tensorflow~\cite{tensorflow2015-whitepaper} implementation of MLP and the tuning of hyperparameters is performed on the training set using the Hyperband tuning algorithm~\cite{li:18}. The tuned hyperparameters include (value range in brackets): number of layers (2 - 5), number of nodes in each layer (64 - 1024), dropout rate (0 - 0.5), and learning rate ($10^{-5}$ - $10^{-2}$). One dropout layer is placed after the dense layers and before the output layer. The tanh activation function is used in all dense layers. For the output layer, we use the sigmoid and the softmax activation for the binary and the multiclass classification, respectively.
The selected hyperparameters for each case are shown in Table~\ref{tab:hyperparameters}.

\begin{table}[t]
\centering
\resizebox{\columnwidth}{!}{
\begin{tabular}{cc|cccc}
Features      & Classif. & Layers & Nodes & DR & LR \\ \hline \hline
Only chemical & Binary   &    5   & {\footnotesize 384,832,256,640,512} & 0.3 & $10^{-4}$\\
Only chemical & Multi    &    4   & 448,1024,896 & 0.4 & $10^{-3}$\\
Fathead Minn. & Binary   &    3   & 192,192,704 & 0.5 & $10^{-2}$\\
Fathead Minn. & Multi    &    3   & 640,128,448 & 0.2 & $10^{-3}$\\
Rainbow Trout & Binary   &    3   & 320,128,896 & 0.1 & $10^{-2}$\\
Rainbow Trout & Multi    &    3   &  960,192,448 & 0.3 & $10^{-3}$\\
All features  & Binary   &    2   & 960,64 & 0.0 & $10^{-4}$\\
All features  & Multi    &    3   & 448,640,256  & 0.1 & $10^{-4}$\\
\end{tabular}
}
\caption{Optimal hyperparameters selected to construct the final MPL for each case as a result from hyperparameter tuning. Hyperparameters are: number of dense layers (Layers), number of nodes in each layer (Nodes), dropout rate (DR), and learning rate (LR).} 
\label{tab:hyperparameters}
\end{table}

\subsection{K Nearest Neighbors}
The KNN algorithm assigns to an unknown entry the target that appears most often among the K nearest-neighbor data points. 
If two values are equally present, the one which is most present in the training set is chosen. For this reason, when the dataset is unbalanced (\textit{e.g.} Tab.~\ref{tab:acc-noneco-mixing}) even values of K give a higher performance.
We use the KNN functions from SciKit Learn~\cite{scikit-learn}.

\begin{figure}
    \centering
    \includegraphics[width = 0.4\textwidth]{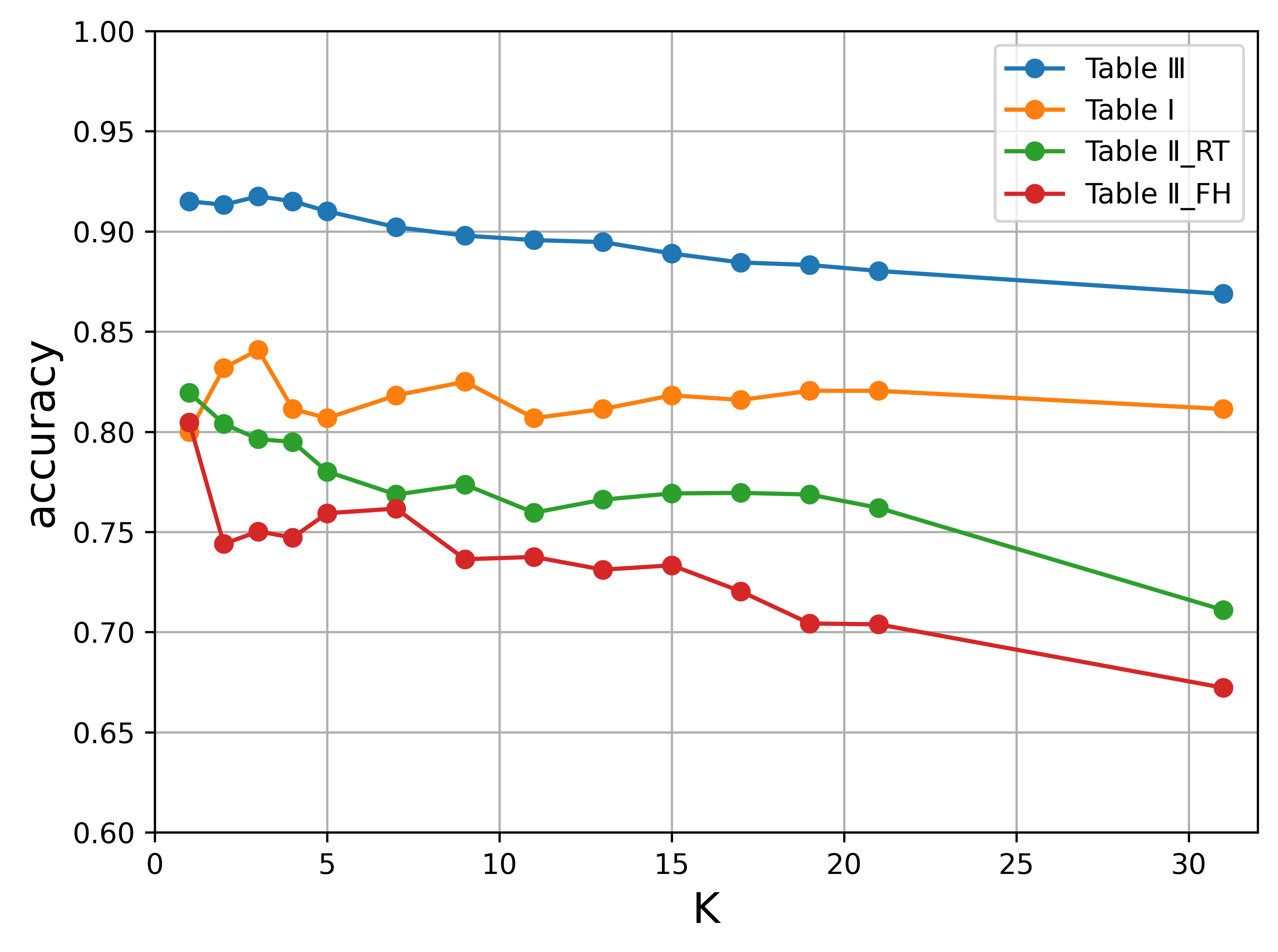}
    \caption{Validation performance of the KNN models from Tab.~\ref{tab:acc-noneco-mixing} (using $\vec x_\mathrm{ch}$ without distinguishing among taxa), Tab.~\ref{tab:acc-noneco-nomixing} (using $\vec x_\mathrm{ch}$, training on a single species (Rainbow Trout or Fathead Minnow) and testing on all the other taxa) and Tab.~\ref{tab:acc-eco} (ecotoxicological dataset $\vec x$) as a function of K. 
    }
    \label{fig:knn-tuning-eco}
\end{figure}

We select and report for the value of K that gives the highest accuracy value. Differently from what is found in Ref.~\cite{hou:20}, we have a better test performance with small values of K. In Fig.~\ref{fig:knn-tuning-eco} we show the validation performance for the different values of K that we tested, for both the \C and the \CTE setup.

\subsection{RASAR}
The RASAR model, presented in Ref.~\cite{luechtefeld:18}, comes in two variants. The S-RASAR requires data on a single effect (mortality in our case), whereas the DF-RASAR integrates several effects. 

\subsubsection{Simple RASAR}
The S-RASAR is composed of two steps. In the first step, a metadataset is created through a nearest-neighbor algorithm, while in the second step, the metadataset is trained through a supervised learning method. The first step does not require any hyperparameter tuning, while the second follows the protocol described in \ref{app:LR+RF}.

For each entry $i$ in the training set, we save the normalized distance $\tilde{d}_+$ from the nearest positive,\footnote{
In Ref.~\cite{luechtefeld:18}, the similarity $s$ with the nearest positive and negative is used. Since the distance $\tilde{d}$ is normalized to 1, it is equivalent to use on or the other, because $s=1-d$.
}
and the one from the nearest negative outcome, $\tilde{d}_-$ (we use the distances defined in Sec.~\ref{sec:models}).
This results in a meta entry 
\begin{equation}
    m_i^\mathrm{(S)}=[\tilde{d}_i^+,\tilde{d}_i^-]
\end{equation}
We now have a supervised learning problem, where we need to deduce the labels ${y_2}_i$ from $m_i$. We did this through LR or RF. 
\footnote{Since we did not remark a difference between LR and RF, we only report the results using RFs, in order to be using the same supervised algorithm both in the S-RASAR and in the DF-RASAR.}

\paragraph*{Multiclass}
In Ref.~\cite{luechtefeld:18}, this algorithm is presented only as a binary classification algorithm, but we can easily extend it to multi-class classification.
It is enough to measure the distance to the nearest element of each class. The multi-class metadata point is then
\begin{equation}
    m_i^\mathrm{(S)}=[\tilde{d}_i^1,\tilde{d}_i^2,\ldots,\tilde{d}_i^{C}]\,,
\end{equation}
where $C$ is the number of classes.

\subsubsection{Data Fusion RASAR}
The DF-RASAR extends the idea of the S-RASAR, by including different sources of information in the metadataset. Since the definition in the binary case is already provided in Ref.~\cite{luechtefeld:18}, we define it here in the case of $C$ classes.

The first, unsupervised, step is performed for $E$ extra effects, beyond the target one, which we denote with a 0 (in our study, mortality).
For each extra effect $e$, we construct an extra metadataset. This metadataset contains $C+1$ elements. The first $C$ elements are the distances to the nearest experiment of each label (distance to the nearest 0, to the nearest 1, $\ldots$), and the $(C+1)^\mathrm{th}$ is the label of effect $e$.
{In other words, if chemical A was tested on taxon B for mortality, now we are including the impact of A on B based on another effect $e$. If this information is unavailable, we use an extra "unknown" label. In the multiclass problem, the introduction of an "unknown" label implies that the six labels are no longer ordered and need to be treated as categorical variables.} 

In formulas, for each extra effect $e$, the metadatapoint corresponding to the $i^\mathrm{th}$ experiment (\textit{i.e.} a taxon-chemical couple) is
\begin{equation}
    \hat{m}_i(e) = [y_C(e), \tilde{d}_i^1(e),\tilde{d}_i^2(e),\ldots,\tilde{d}_i^{C}(e)]\,,
\end{equation}
where $y_C(e)$ is the $C$-class label (see \ref{app:multiclass}) of the $e^\mathrm{th}$ effect.

When the features used for the $e^\mathrm{th}$ effect are different from those of the target effect. In those cases, we restrict the distance to the common features.

As a result, the $i^\mathrm{th}$ meta datapoint for a $C$-class DF-RASAR takes the form
\begin{align}
m_i^\mathrm{(DF)} =& \left(m_i^\mathrm{(S)},\hat{m}_i(1),\ldots,\hat{m}_i(E)\right)=\\
=&\begin{bmatrix}
\begin{pmatrix}
\tilde{d}_i^1(0)\\\vdots\\\tilde{d}_i^C(0)\\-
\end{pmatrix}
\begin{pmatrix}
\tilde{d}_i^1(1)\\\vdots\\\tilde{d}_i^C(1)\\{y_C}_i(1)
\end{pmatrix}
\ldots
\begin{pmatrix}
\tilde{d}_i^1(E)\\\vdots\\\tilde{d}_i^C(E)\\{y_C}_i(E)
\end{pmatrix}
\end{bmatrix}\,.
\end{align}
We then use RFs on this metadataset to relate it to the target, ${y_C}_i(0)$.

\section{Performance metrics} \label{app:metrics}
Here, we explicitly define the perofmance metrics used in this manuscript: accuracy ($a$), recall ($r$, often called sensitivity), specificity ($s$), precision ($p$) and F1 score (F$_1$).

The accuracy is the most intuitive of these quantities, and measures the fraction of correct answers,
\begin{equation}
    a = \frac{\# \text{correct guesses}}{\text{total}}\,.
\end{equation}
In terms of the number of true positives ($TP$), true negatives ($TN$), false positives ($FP$) and false negatives ($FN$), the remaining metrics are defined as
\begin{align}
    % a  =&~ \frac{TP+TN}{TP+TN+FP+FN}\,, \\[1ex]
    r  =&~ \frac{TP}{TP+FN}\,,\\[1ex]
    s  =&~  \frac{TN}{TN+FP}\,,\\[1ex]
    p  =&~  \frac{TP}{TP+FP}\,,\\[1ex]
    \text{F}_1 =&~ 2\frac{rp}{r+p}\,. 
\end{align}
The recall measures, out of all the positives in the test set, how many of those were guessed correctly. The specificity does the same, but for the negatives. In balanced datasets (\textit{i.e.} with a similar number of positives and negatives), these are usually similar. 
With unbalanced datasets, high accuracies can be an artifact of the unbalance, and this usually results in very different values of $r$ and $s$. In those cases, a better indicator of the performance is either the balanced accuracy ($b=\frac{r+s}{2}$), or the F$_1$ score {~\cite{luque:19}}. 

For multiclass classification, we calculate the metrics on each class, and then average between the classes without taking into account how many examples there are in each class (this is called a macro average~\cite{opitz:19}).
In the multiclass case, the sensitivity does not make too much sense, since the negatives of one class include several classes. Therefore, we report the precision, which indicates, out of all the examples that were labeled as positive (for a given class) by the classifier, how many of those were actually correct.

\section{Reproducibility of the experiments: formal definitions}\label{app:formal-definitions}
Here, we define formally the quantities $A_\mathrm{up}, R_\mathrm{up}, S_\mathrm{up}$ and $\text{BA}_\mathrm{up}$, that we introduced in Sec.~\ref{sec:repeated-exp}.
Let $y(i)=0,\ldots,C-1$ be a generic label (in this work we have $C$ set to 2 or 5 classes), related to an experiment $i$, which is performed $n_i$ times. Let us call $n_\mathrm{rep}$ the number of experiments that were performed at least twice, and $y_j(i)$ the label resulting from the $j^\mathrm{th}$ experiment. Let us also call $\yi$ the label of experiment $i$ that appears the most times, over the $n_i$ times that the experiment is performed. Then,
we have
\begin{equation}\label{eq:aup}
    A_\mathrm{up} \equiv 
    \frac{1}{n_\mathrm{rep}}\displaystyle\sum_{i=1}^{n_\mathrm{rep}} \left[
    \frac{\sum_{j=1}^{n_i}
    \delta\left(y_j(i), \yi\right)
    }{n_i}\right]\,,
\end{equation}
where $\delta(a,b)$ is the Kronecker function, which is equal to 1 if $a=b$, and equal to 0 if $a\neq b$. Therefore, the term in square brackets is the fraction of evaluations of the experiment that resulted in the most common outcome.

By restricting the measurement of $A_\mathrm{up}$ to the experiments with a majority of positive outcomes, we get an upper bound to the recall, $R_\mathrm{up}$. By restricting to the negative outcomes, we obtain an upper bound to the specificity, $S_\mathrm{up}$: 
\begin{align}\label{eq:rup}
    R_\mathrm{up} &
    \equiv 
    \frac{\displaystyle\sum_{i=1}^{n_\mathrm{rep}} \left[
    \frac{\delta\left(1, \yi\right)}{n_i}\sum_{j=1}^{n_i}
    \delta\left(y_j(i), \yi\right)
    \right]}{\displaystyle\sum_{i=1}^{n_\mathrm{rep}}\delta\left(1, \yi\right)},\\\label{eq:sup}
    S_\mathrm{up}&
    \equiv 
    \frac{\displaystyle\sum_{i=1}^{n_\mathrm{rep}} \left[
    \frac{\delta\left(0, \yi\right)}{n_i}\sum_{j=1}^{n_i}
    \delta\left(y_j(i), \yi\right)
    \right]}{\displaystyle\sum_{i=1}^{n_\mathrm{rep}}\delta\left(0, \yi\right)}.
\end{align}

Here, are separating the experiments in two subsets. In Eq.~\eqref{eq:rup} we are only considering the experiments with a majority of $y_j(i)=1$ outcomes. This is enforced by the term $\delta\left(1, \yi\right)$, which is zero unless the label is a 1. Since we discarded a fraction of the experiments, the normalization factor is no longer $n_\mathrm{rep}$, but the total number of experiments with a majority of $y_j(i)=1$ outcomes.
The definition in Eq.~\eqref{eq:sup} is equivalent, but we consider the experiments with a majority of $y_j(i)=0$. 

We obtain an estimated upper bound for the balanced accuracy by taking 
\begin{equation}
    \text{BA}_\mathrm{up}\equiv\frac{R_\mathrm{up}+S_\mathrm{up}}2. 
\end{equation}

For multiclass classification, $A_\mathrm{up}$ is calculated in the same manner.
Since in this case the specificity is not informative, we restrict to the recall.
We calculate separately the average recall $R_\mathrm{up,c}$ related to each class $c$, and then average over all the classes (\textit{i.e.} we measure the macro-averaged recall):
\begin{equation}
    R_\mathrm{up}^\mathrm{(MC)} = \frac{1}{C}\sum_{c=0}^{C-1}R_\mathrm{up,c}\,.
\end{equation}

\section{Permutation-based feature importance}\label{app:importance}
Since the impurity-based feature importance analysis is at risk of giving too much weight to high-cardinality features~\cite{altmann:10}, we also used permutation importance, to ensure that the species (which indeed has a high cardinality), indeed is one of the most important features.
\begin{figure}[t]
    \centering
    \includegraphics[width=\columnwidth]{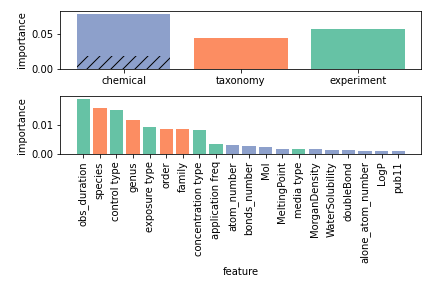}
    \caption{Permutation-based feature importance from our best RF model. \textbf{Top}:  Every bar corresponds to a collection of features. The bottom (hatched) part of the bar corresponding to the chemicals is related to the chemical fingerprints, whereas the top part depicts the importance of the pubchems. \textbf{Bottom}: feature importance of the single features. For visualization purposes, we only include the 20 most important features.}
    \label{fig:feature_importance_permutation}
\end{figure}
In the bottom set of Fig.~\ref{fig:feature_importance_permutation} we see that, also when using permutation importance, the species remains one of the most important features.
We notice that the importance of chemical-related features is now diminished with respect to the impurity-based method (this is also visible from the top chart). This can be expected, since the features in $\vec x_\mathrm{ch}$ are not linearly independent (\textit{i.e.} similar information can be carried by more than one feature).

\section{Multi-class performances} \label{app:multiclass}
In this section, we show the results of the main paper, but with a multi-class target,
\begin{equation}\label{eq:multiclass}
    y_5 = \begin{cases}
    4 ~~~\text{if~~}                      & c~\leq 0.1~\text{mg}/\ell \,,\\
    3 ~~~\text{if~~} 0.1~\text{mg}/\ell~ <& c~\leq 1~~~\text{mg}/\ell \,,\\
    2 ~~~\text{if~~} 1~~~\text{mg}/\ell~ <& c~\leq10~\,\text{mg}/\ell \,,\\
    1 ~~~\text{if~~} 10~\,\text{mg}/\ell~<& c~\leq100\text{mg}/\ell \,,\\
    0 ~~~\text{if~~} 100\text{mg}/\ell  ~<& c~\,.
    \end{cases}
\end{equation}

The binary and multi-class (\ref{app:multiclass}) thresholds are summarized in Tab.~\ref{tab:classes}.

\begin{table}[th]
\centering
\resizebox{\columnwidth}{!}{
\begin{tabular}{c|c|c|c}
LC50 interval (mg/$\ell$)  & Binary Score & Multiclass Score \\ \hline \hline
($-\infty$,  $10^{-1}$] &\cellcolor[HTML]{ff9797} 1 & \cellcolor[HTML]{ff9797} 4 \\
($10^{-1}$,  $10^{0}$] &\cellcolor[HTML]{ff9797} 1 &\cellcolor[HTML]{ffad7f} 3 \\
($10^{0}$,  $10^{1}$] &\cellcolor[HTML]{9dfbb6} 0 &\cellcolor[HTML]{edc977} 2 \\
($10^{1}$,  $10^{2}$] &\cellcolor[HTML]{9dfbb6} 0 &\cellcolor[HTML]{cbe489} 1 \\
($10^{2}$,  $+\infty$) &\cellcolor[HTML]{9dfbb6} 0 & \cellcolor[HTML]{9dfbb6} 0 \\
\hline
\end{tabular}
}
\caption{LC50 intervals for the binary and multi-class labeling. For the binary labeling, we can say that the labels distinguish more (0) from less (1) toxic outcomes.
For the multi-class, we can refer to the standard thresholds~\cite{oecd:02,li:17b}:
% Non toxic (0),
non-toxic (0), % li:17b
% hazardous (1), 
slightly toxic (1), % li:17b
% toxic (2), 
moderately toxic (2), % li:17b
%very toxic (3), 
highly toxic (3), % li:17b
% extremely toxic (4).
very highly toxic (4). % li:17b
}
\label{tab:classes}
\end{table}

In Tabs.~\ref{tab:acc-noneco-mixing-multiclass}, \ref{tab:acc-noneco-nomixing-multiclass}, \ref{tab:acc-eco-multiclass} and \ref{tab:alphas-multiclass} 
we report, for multi-class classification, the same results that in the main text we reported for binary classification (Tabs.~\ref{tab:acc-noneco-mixing}, \ref{tab:acc-noneco-nomixing}, \ref{tab:acc-eco} and \ref{tab:alphas}, where we report the precision instead of the specificity, since it is better defined for multi-class classification).
We observe the same trend shown for binary classification. The F1-score increase obtained by including taxonomy and experiment-related information is of around 15\%.
The best accuracy that we are able to reach with multi-class classification is 0.781 (Tab.~\ref{tab:alphas-multiclass}), which is close to the estimators of the reproducibility of experiments ($A_\mathrm{up}=0.878$, Tab.~\ref{tab:exp-acc}), indicating that likely our models are close to achieving the best possible performance in this dataset. If we rescale the maximum obtained accuracy by $A_\mathrm{up}$, we obtain $0.781/0.878\simeq0.890\approx89\%$, which is close to we obtained for binary classification. The recall is even closer to its estimate maximum $R_\mathrm{up}$, as it is $0.787/0.833\simeq0.945\approx95\%$.

Even in the multi-class case, the best models are RF, MLP and DF-RASAR, as it was for binary classification.
From the values of the $\alpha$ parameters in Tab.~\ref{tab:alphas} we find the same indications, with $\alpha_\mathrm{cat}$ being smaller, though not negligibly, than $\alpha_\mathrm{Euc}$ and $\alpha_\mathrm{pubc}$, 
and the pubchems seemingly less important when a smaller number of neighbors is taken into account.

\begin{table}[!htb]
\centering
\resizebox{\columnwidth}{!}{
\begin{tabular}{c||c|c|c|c}
Model       & Accuracy  & Recall    & Precision & F$_1$        \\ \hline \hline
LR          & 0.45(1) & 0.400(9)  & 0.45(3) & 0.39(1) \\
1-NN        & 0.459(9)  & 0.459(5)  & 0.461(8)  & 0.458(6)  \\
% 7-NN  & 0.461(6)  & 0.421(6)  & 0.481(11) & 0.431(7)  \\
RF          & 0.545(9)  & 0.517(7)  & 0.55(1) & \textbf{0.525(8)}  \\
MLP         & 0.47(3) & 0.47(4) & 0.48(2) & 0.47(5) \\
S-RASAR(RF) & 0.445(3)  & 0.422(7)  & 0.45(2) & 0.426(8)  \\
DF-RASAR    & 0.48(2) & 0.46(2) & 0.48(2) & 0.46(2)
\end{tabular}
}
\caption{
Metrics of different multi-class models, trained and tested on $\vec x_\mathrm{ch}$, \textit{i.e.} without distinguishing among taxa and experimental details (same as Tab.~\ref{tab:acc-noneco-mixing}, but for multi-class classification.).
The numbers in parentheses are an uncertainty estimate (see Sec.~\ref{sec:errors}).
}
\label{tab:acc-noneco-mixing-multiclass}
\end{table}

\begin{table}[!htb]
\centering
\resizebox{\columnwidth}{!}{
\begin{tabular}{c||c|c|c|c|c}
Model    & Training Taxon   & Acc.           & Recall         & Prec.          & F$_1$             \\ \hline
1-NN     & Fathead   Minnow & 0.493          & 0.497          & 0.503          & 0.498          \\
RF       & Fathead   Minnow & {0.512} & {0.518} & {0.525} & \textbf{0.518} \\
MLP      & Fathead   Minnow & 0.461          & 0.449          & 0.478          & 0.458          \\
S-RASAR  & Fathead   Minnow & 0.481          & 0.487          & 0.502          & 0.482          \\
DF-RASAR & Fathead   Minnow & 0.498          & 0.505          & 0.515          & 0.501          \\
1-NN     & Rainbow   Trout  & 0.523          & 0.532          & 0.532          & 0.525          \\
RF       & Rainbow   Trout  & {0.541} & {0.545} & {0.553} & \textbf{0.541} \\
MLP      & Rainbow   Trout  & 0.473          & 0.500          & 0.475          & 0.477          \\
S-RASAR  & Rainbow   Trout  & 0.493          & 0.494          & 0.522          & 0.487          \\
DF-RASAR & Rainbow   Trout  & 0.523          & 0.526          & 0.539          & 0.522         
\end{tabular}
}
\caption{
Multi-class out-of-domain generalization, for different models,  using  only $\vec x_\mathrm{ch}$ (same as Tab.~\ref{tab:acc-noneco-nomixing}, but for multi-class classification).   Here,  models  are  trained  on  data  coming from  a  single  taxon  (Rainbow  Trout  or  Fathead  Minnow), and  they  are  tested  on  all  the  other  taxa  contained  in  the data set.
}
\label{tab:acc-noneco-nomixing-multiclass}
\end{table}

\begin{table}[!htb]
\centering
\resizebox{\columnwidth}{!}{
\begin{tabular}{c||c|c|c|c}
Model       & Accuracy  & Recall    & Precision & F$_1$        \\ \hline \hline
LR          & 0.639(5)  & 0.643(5)  & 0.641(6)  & 0.642(5)  \\
1-NN        & 0.760(3)  & 0.765(3)  & 0.763(3)  & 0.764(3)  \\
RF          & \textbf{0.780(2)}  & \textbf{0.787(2)}  & \textbf{0.784(2)}  & \textbf{0.785(2)}  \\
MLP         & 0.78(1) & 0.78(1) & 0.78(1) & 0.78(1) \\
S-RASAR     & 0.745(3)  & 0.750(2)  & 0.748(3)  & 0.749(2)  \\
DF-RASAR    & 0.778(2)  & 0.784(2)  & 0.781(2)  & 0.782(2) 
\end{tabular}
}
\caption{Performance of our models, trained on the full \CTE feature space, in multi-class classification (same as Tab.~\ref{tab:acc-eco}, but for multi-class).
}
\label{tab:acc-eco-multiclass}
\end{table}

\begin{table}[!htb]
\centering
\resizebox{\columnwidth}{!}{
\begin{tabular}{c||c|c|c|c}
Model    & Accuracy & Recall   & Precision & F$_1$       \\ \hline
1-NN     & {0.769(1)} & {0.775(1)} & {0.775(1)}    & {0.774(1)} \\
3-NN     & 0.751(4) & 0.756(3) & 0.753(4)    & 0.754(3) \\
5-NN     & 0.728(4) & 0.729(4) & 0.729(4)    & 0.729(4) \\
S-RASAR  & 0.770(2) & 0.775(3) & 0.773(2)    & 0.774(2) \\
DF-RASAR & \textbf{0.781(3)} & \textbf{0.784(2)} & \textbf{0.783(3)}    & \textbf{0.783(2)}
\end{tabular}
}
\caption{Estimates of $y_2$ from the whole dataset $\vec x= (\vec x_\mathrm{ch}, \vec x_\mathrm{tax}, \vec x_\mathrm{ex})$, using the distances defined in Eq.~\eqref{eq:dist} (same as Tab.~\ref{tab:alphas}, but for multi-class).
}
\label{tab:alphas-multiclass}
\end{table}

\begin{table}[!htb]
\centering
\begin{tabular}{c|ccc}
Model    & $\alpha_\mathrm{Euc}$ & $\alpha_\mathrm{pubc}$ & $\alpha_\mathrm{cat}$ \\ \hline
1-NN     & 1                     & 0.3562                 & 0.0127                \\
3-NN     & 1                     & 4.8939                 & 0.0853                \\
5-NN     & 1                     & 3.0392                 & 0.0672                \\
S-RASAR  & 1                     & 0.3562                 & 0.0127                \\
DF-RASAR & 1                     & 0.3562                 & 0.0127               
\end{tabular}
\caption{Values of $\alpha_\mathrm{Euc}$, $\alpha_\mathrm{pubc}$ and $\alpha_\mathrm{cat}$ [Eq.~\eqref{eq:dist}] for different models for multiclass.
}
\label{tab:alpha-values-multiclass}
\end{table}

\section{Stratified splitting} \label{app:stratified}
Here, we show how our models perform in the case of a different splitting of the dataset. We choose two ways of doing stratified splitting: by chemical and by taxon. In the first, each CV fold, as well as the hold-out test set, contain exclusively different chemicals. In the second, they contain different taxa.

To do this, we used the GroupShuffleSplit function from SciKit Learn~\cite{scikit-learn} to split the train and test dataset, ensuring the (chemical, taxon) groups in the test dataset are independent of those in the training dataset. In the 5-fold CV, we used the GroupKFold function. 
For the LR and RF we adopt mitigate class imbalance through class reweighting.

\subsection{Splitting by chemical}

In Table~\ref{tab:acc-noneco-mixing-c} we show the performance of the \C approach when stratifying the performances by chemical, and in Table~\ref{tab:acc-eco-c} we show the performance of the \CTE approach in the same conditions. The performances of the \CTE approach are better than the \C approach also in this case.

\begin{table}[!htb]
\centering
\resizebox{\columnwidth}{!}
{
\begin{tabular}{c||c|c|c|c}
Model    & Accuracy  & Recall    & Specificity & F$_1$        \\ \hline \hline

LR       & 0.810(8) & 0.278(26) & 0.946(6) & 0.373(27) \\
2-NN     & 0.820(8) & 0.209(13) & 0.969(6) & 0.313(16) \\
RF       & 0.818(7) & 0.385(20) & 0.931(7) & 0.467(9)  \\
S-RASAR  & 0.843(6) & 0.261(7)  & 0.989(4) & 0.400(9)  \\
DF-RASAR & 0.870(3) & 0.338(16) & 0.983(6) & 0.477(14)

\end{tabular}
}
\caption{Estimates of $y_2$ from the whole dataset with chemical stratification split using the \C approach, $\vec x= (\vec x_\mathrm{ch})$.
}
\label{tab:acc-noneco-mixing-c}
\end{table}

\begin{table}[!htb]
\centering
\resizebox{\columnwidth}{!}
{
\begin{tabular}{c||c|c|c|c}
Model    & Accuracy  & Recall    & Specificity & F$_1$        \\ \hline \hline
LR       & 0.789(23) & 0.924(22) & 0.590(47)   & 0.693(40) \\
3-NN     & 0.823(24) & 0.751(35) & 0.878(19)   & 0.785(42) \\
RF       & 0.764(21) & 0.799(3)  & 0.699(48)   & 0.671(42) \\
S-RASAR  & 0.751(15) & 0.867(11) & 0.560(36)   & 0.629(39) \\
DF-RASAR & 0.730(16) & 0.948(24) & 0.670(37)   & 0.671(26)
\end{tabular}
}
\caption{Accuracy on binary classification with dataset stratified splitting by chemical stratification split using the \CTE approach, $\vec x= (\vec x_\mathrm{ch}, \vec x_\mathrm{tax}, \vec x_\mathrm{ex})$.
}
\label{tab:acc-eco-c}
\end{table}

\subsection{Splitting by species}
In Table~\ref{tab:acc-eco-t} we show the performance of the CTE approach when stratifying the performances by species.

\begin{table}[!htb]
\centering
\resizebox{\columnwidth}{!}
{
\begin{tabular}{c||c|c|c|c}
Model    & Accuracy  & Recall    & Specificity & F$_1$     \\ \hline \hline
LR       & 0.852(10) & 0.923(13) & 0.678(10)   & 0.727(19) \\
9-NN     & 0.840(5)  & 0.774(24) & 0.885(12)   & 0.797(18) \\
RF       & 0.895(10) & 0.940(14) & 0.834(21)   & 0.871(5)  \\
S-RASAR  & 0.852(10) & 0.927(10) & 0.737(26)   & 0.797(19) \\
DF-RASAR & 0.885(7)  & 0.964(12) & 0.842(46)   & 0.861(39)
\end{tabular}
}
\caption{Accuracy on binary classification with dataset stratified splitting by species stratification split using the \CTE approach, $\vec x= (\vec x_\mathrm{ch}, \vec x_\mathrm{tax}, \vec x_\mathrm{ex})$.}
\label{tab:acc-eco-t}
\end{table}

\bibliographystyle{elsarticle-num}
\bibliography{marco}

\begin{thebibliography}{10}
\bibitem{mltox-code}
We provide our code at \texttt{github.com/mbaityje/ML-Tox}.
\bibitem{oecd:02}
{Organisation for Economic Co-operation and Development}, Guidance Document on
  the Use of the Harmonised System for the Classification of Chemicals which
  are Hazardous for the Aquatic Environment, OECD Publishing, 2002.
\end{thebibliography}

\end{document}